\documentclass[aps,preprint,nofootinbib]{revtex4-1}
\pdfoutput=1
\usepackage{physics,amsmath,amssymb,hyperref,color}

\begin{document}
\count\footins = 1000
\title{Path Integral for Mixed Tunneling, Polychronic Tunneling and Quantum Gravity
}
\newcommand{\bfx}{{\bf x}}
\newcommand{\bfy}{{\bf y}}

\renewcommand{\theequation}{\thesection.\arabic{equation}}

\author{Yutaro Shoji}
\affiliation{Racah Institute of Physics, Hebrew University of Jerusalem, Jerusalem 91904, Israel}

\begin{abstract}
    Quantum tunneling in a many-body system is much more non-trivial than that in a one-body system. The most characteristic phenomenon is the mixed tunneling, which has been studied in many fields for decades.
    For instance, let us consider a system where there are two coupled particles and only one of them feels a potential barrier. Quantum tunneling of such a system is not described by either Euclidean or Lorentzian time evolution and the exponent of the WKB wave function becomes complex.
    Recently, a similar phenomenon, polychronic tunneling, has been proposed in quantum gravity, which enhances the decay rate of a meta-stable vacuum by many orders of magnitude.
    In this paper, we present path integral formalism that is applicable to such systems. The formalism can be directly extended to quantum gravity and has some implications on the problem of time in quantum gravity.
    We also discuss a possible relation to the conventional path integral.
\end{abstract}

\maketitle

\section{Introduction}
Quantum theory of a many-body system is a foundation of modern science and has been successful in various fields such as quantum chemistry, condensed matter physics, nuclear physics and particle physics. The most remarkable successes are the precise prediction of the muon magnetic moment \cite{Aoyama:2020ynm} and that of the collider results \cite{ATLAS:2022djm,CMS:web}. The theoretical predictions and the experimental results agree over a very wide energy range, demonstrating the validity of quantum field theory.

Although theories of quantum many-body systems have been extensively tested and are nearly established, there still remain subtleties.
One of these is the understanding of the so-called mixed tunneling phenomenon. The simplest example is quantum tunneling of two coupled particles, one of which feels a potential barrier and the other does not.
In such a case, while one particle is tunneling, the other evolves in real time.
Thus, neither Lorentzian time nor Euclidean time can evolve the entire system and this results in a complex exponent of the WKB wave function \cite{PhysRevA.41.32,doi:10.1063/1.466899}.
Another example is a tunneling process in which the barrier hight is slightly lower than the kinetic energy. In such a system, the interaction with the potential excites oscillators and prevents the classical transition dynamically. This phenomenon is called the dynamical tunneling and has been predicted for a long time ago \cite{doi:10.1063/1.441832,CALDEIRA1983374,Tomsovic_2001}. In a strong coupling regime, the change of quantum numbers during the tunneling process becomes non-negligible \cite{Bezrukov:2003tg} and one needs to evaluate the complex exponent.
In addition to these examples, the mixed tunneling has been observed and discussed in various systems such as chemical reactions \cite{doi:10.1063/1.466899,doi:10.1063/1.468526,doi:10.1063/1.471029}, ultracold atoms \cite{834911}, molecular collisions \cite{doi:10.1063/1.1674535,doi:10.1063/1.1678610}, and nuclear fusion \cite{Tanimura:1985hnh,Takigawa:1993ha,Balantekin:1997yh}.

The estimation of the complex exponent is, however, a very non-trivial task even though in principle such a solution can be obtained by directly solving the Shr\"odinger equation.
For example, in the conventional path integral formalism, we integrate either $e^{iS}$ or $e^{-S_E}$ over the possible paths connecting two states, where $S$ is a Lorentzian action and $S_E$ is a Euclidean action.
By definition, it does not generate a complex exponent at the leading order in $\hbar$ and the complexity is only visible after evaluating loop integrals with somewhat ad-hoc analytic continuation followed by the dilute gas approximation \cite{Callan:1977pt}. However, the validity of the analytic continuation is unclear \cite{Andreassen:2016cvx} and the non-perturbative definition of analytic continuation seems to be lacking.

Several methods to deal with the mixed tunneling have been proposed in different fields. One is called the adiabatic approximation \cite{Tanimura:1985hnh,Takigawa:1993ha}. It relies on the adiabatic theorem \cite{Born} of quantum mechanics, {\it i.e.} a state remains in its instantaneous eigenstate if the perturbing potential is either adiabatic or diabatic. It means that if the timescales of the oscillators coupled to the tunneling particle are significantly different from that of the tunneling, we can ignore the change of the state due to the oscillators during the tunneling. The same is true if the couplings to the oscillators are very weak. However, this method is limited to adiabatic, diabatic and weak-coupling cases, and the time scale of quantum tunneling itself has been one of the controversial issues in quantum mechanics \cite{Wigner:1955zz,Hauge:1989zz,Landauer:1994zz,Davies:2004fty,Hartman:1962jlq,WINFUL20061,Sokolovski2021}.

There are also methods based on the Huygens-Fresnel principle \cite{PhysRevA.41.32,doi:10.1063/1.466899,doi:10.1063/1.468526,doi:10.1063/1.3490087}. A complex action is obtained by solving the complex Hamilton-Jacobi equation assuming the Huygens-type wave propagation. They construct wave fronts sequentially for given boundary conditions and integrate the Hamilton-Jacobi equation towards the wave normals. Although it is a very versatile method, the numerical procedure is not easy even in a two-dimensional system. Recently, a similar construction was proposed in \cite{Oshita:2021aux}, where quantum tunneling in quantum gravity is analyzed in the WKB approximation. In their formulation, wave fronts are determined almost uniquely thanks to the additional constraints on the system: the Hamiltonian constraint and the momentum constraints.
The method presented in this paper is hinted by \cite{Oshita:2021aux} and offers more numerically friendly formulas than \cite{PhysRevA.41.32,doi:10.1063/1.466899,doi:10.1063/1.468526,doi:10.1063/1.3490087}.

Another popular method is called the complex trajectories, which has been proposed originally to estimate an instanton-mediated $2\to\rm any$ transition cross section in a high energy collision \cite{Rubakov:1991fb,Rubakov:1992ec}. It is based on the conjecture that the cross section of $2\to\rm any$ can be reproduced by that of $n\to\rm any$ in an appropriate limit. The amplitude of $n\to\rm any$ is calculated by the double analytic continuation, where both time and the fields are continued to the complex plane. It allows us to use a semi-classical technique to estimate the complex exponent. Although the feasibility of the method has been demonstrated by comparison with the instanton result \cite{Tinyakov:1991fn,Mueller:1992sc}, there is no rigorous proof available in the literature.
We will not go into it further since the theoretical comparison against our method is highly non-trivial.

In this paper, we propose an alternative path integral formalism that can treat mixed tunneling explicitly. We construct the path integral formula from the local energy conservation law, instead of the Schr\"odinger equation for the whole system.
It is actually the same starting point as quantum gravity, where there is no naive Schr\"odinger equation and the system is described by the local constraints called the Hamiltonian constraint and the momentum constraints.
Once we impose the local energy conservation law, we cannot rely on the standard procedure to derive path integral since the global time variable disappears from the theory. This is known as the problem of time in quantum gravity (see \cite{Isham:1992ms,Anderson:2010xm} for reviews). We overcome this problem and obtain path integral formulas, which have square root actions.
They are identical to the conventional path integral when the entire evolution is exclusively Euclidean or Lorentzian. However, our path integral formulas accommodate the coexistence of Euclidean and Lorentzian evolution and can describe the mixed tunneling.

Since the mixed tunneling is a general feature of a quantum many-body system, a similar process should exist in quantum field theory. 
Recently, such a process, polychronic tunneling, has been proposed in quantum gravity \cite{Oshita:2021aux}, where Euclidean and Lorentzian domains coexist and the exponent of the WKB wave function becomes complex. 
It is thus different from the traditional quantum tunneling in quantum field theory \cite{Banks:1973ps,Coleman:1977py,Callan:1977pt,Coleman:1980aw}, where the tunneling is described in Euclidean spacetime and the exponent is given by a Euclidean action.
It was found that the polychronic tunneling tends to give a much higher tunneling rate than the conventional rate and thus has a disastrous effect on the stability of a false vacuum.
They also demonstrated numerically that the polychronic tunneling can happen in the decoupling regime of gravity.
Our formalism gives a non-gravitational formulation of the polychronic tunneling and supports their arguments.

Next, we apply our path integral formalism to quantum gravity. We find the local energy conservation law and supplemental equations employed in our formalism have the same structure as that of the Hamiltonian constraint and the momentum constraints in quantum gravity. In fact, with an appropriate choice of auxiliary fields, we see that these path integral formulas agree exactly in the decoupling limit of gravity. It will be beneficial for the understanding of the problem of time since the formulation is parallel to those of non-gravitational cases and the problem is more traceable. In our formalism, time appears as a parameter to evolve operators, which is similar to the emergent time discussed in \cite{York:1972sj,Isham:1995vt,DeWitt:1967yk,Hajicek:1994py}, although we do not employ the WKB approximation.

Finally, we discuss the relation between our path integral formula and the conventional one. With a simple example, we show that the our path integral formula can be obtained from the conventional one by changing the integration variables and performing the integrals over the time dilatations.

This paper is organized as follows. In the next section, we analyze a separable system and give an overview of our formulation. In Section \ref{sec_non_separable}, we find the local energy conservation law for a non-separable system and obtain the path integral formula. Then, it is promoted to quantum field theory in Section \ref{sec_qft}. The path integral formulation of quantum gravity is discussed in Section \ref{sec_gravity}. Then, in Section \ref{sec_relation}, we discuss the relation between our path integral formula and the conventional one. Finally, we give a summary and discussions in Section \ref{sec_summary}.

\section{Separable System}\label{sec_separable}
\setcounter{equation}{0}
We start with a simple example to explain our path integral formalism.
We consider a separable system of $(2N+1)$ particles with the Hamiltonian given by
\begin{equation}
    H=\sum_{a=-N}^N\frac{p_a^2}{2m}+\delta_{a0}V(x^a),
\end{equation}
where only $a=0$ has a potential,
\begin{equation}
    V(x)=
    \begin{cases}
        V_0&(0<x<1)\\
        0&{\rm (otherwise)}
    \end{cases}.\label{eq_separable_potential}
\end{equation}
Here, we consider the same mass for all the particles to see the connection with quantum field theory. 
We assume all the particles have the same initial energy, $0<\mathcal E<V_0$, and calculate the probability to find that all the particles are above one, all starting below zero.

\subsection{Schr\"odinger Equation and Local Energy Conservation Law}\label{subsec_lecl}
It is instructive to see the wrong calculation of the tunneling probability first.
Since the initial state is an energy eigenstate, we consider solving the following time-independent Schr\"odinger equation\footnote{Although we do not consider interaction between particles in this section, we treat the system as a non-separable system so that the formalism can be generalized to a non-separable case. Then, the only equation we have is the Schr\"odinger equation for all the particles.};
\begin{equation}
    \hat H\psi=E\psi,
\end{equation}
with
\begin{equation}
    E=(2N+1)\mathcal E.
\end{equation}
Then, we expand the solution with respect to $\hbar$ as
\begin{equation}
    \psi=\exp[\frac{i}{\hbar}\Theta^{(0)}+\Theta^{(1)}+\dots].
\end{equation}
At the leading order of the WKB approximation, the Schr\"odinger equation becomes
\begin{equation}
    \frac{1}{2m}\sum_{a=-N}^N\qty(\pdv{\Theta^{(0)}}{x^a})^2=E-V(x^0).\label{eq_wkb_wo_grav}
\end{equation}

Since this is the Hamilton-Jacobi equation, a solution can be constructed by the method of characteristics as \cite{Banks:1973ps,Coleman:1977py}
\begin{align}
    \Theta^{(0)}(s_f)-\Theta^{(0)}(s_i)&=\int^{s_f}_{s_i}\dd{s}\sqrt{2m(E-V(x^0))}\sqrt{\sum_{a=-N}^N\qty(\dv{x^a}{s})^2},\label{eq_wrong_result}
\end{align}
where $s$ parameterizes the trajectory of the particles and $\Theta^{(0)}(s)=\Theta^{(0)}(x^{-N}(s),\dots,x^N(s))$. For a review of the method, see Appendix \ref{apx_char}.
This implies, however, that the tunneling probability is one if $E>V_0$. This is contradictory since we know one particle needs to tunnel.
This demonstrates that the Hamilton-Jacobi equation is non-trivial in general and that there are cases where Eq.~\eqref{eq_wrong_result} does not give a correct result, which is indeed commented on in \cite{Banks:1973ps}.

Aside from the time-independent Schr\"odinger equation for all the particles, we have separated equations,
\begin{equation}
    \hat {\mathcal H}_a\psi=0,\label{eq_sch_for_each}
\end{equation}
with
\begin{align}
    \hat{\mathcal H}_a&=\frac{\hat p_a^2}{2m}+\delta_{a0}V(\hat x^a)-\mathcal E.
\end{align}
We could solve these separately and obtain
\begin{align}
    \Theta^{(0)}(s_f)-\Theta^{(0)}(s_i)&=\sum_{a=-N}^N\int_{s_i}^{s_f}\dd{s}\sqrt{2m(\mathcal E-\delta_{a0}V(x^a))}\sqrt{\qty(\dv{x^a}{s})^2}.\label{eq_correct_result}
\end{align}
It gives the correct tunneling rate of
\begin{equation}
    P=\frac{|\psi(s_f)|^2}{|\psi(s_i)|^2}\simeq\exp[-\frac{2}{\hbar}\int_{0}^{1}\dd{x}\sqrt{2m(V_0-\mathcal E)}].\label{eq_correct_rate}
\end{equation}

Even though Eqs.~\eqref{eq_correct_result} and \eqref{eq_wrong_result} look different, they agree with each other in some cases.
For $x<0$ and $x>1$, we can see they give the same result, using the Lorentzian energy conservation for each particle,
\begin{equation}
    \frac{m}{2}\qty(\dv{x^a}{t})^2+\delta_{a0}V(x^a)=\mathcal E,\label{eq_sep_each}
\end{equation}
identifying $t=s$.
However, for $0<x<1$, they disagree because one of the particles does not satisfy the above equation, but Eq.~\eqref{eq_sep_each} with $t=-is$.

For $0<x<1$, we find that the left-hand side of Eq.~\eqref{eq_wkb_wo_grav} is not positive definite and the exponent becomes complex. Following \cite{PhysRevA.41.32,doi:10.1063/1.466899,doi:10.1063/1.468526,doi:10.1063/1.3490087}, let us decompose the exponent as
\begin{equation}
    \Theta^{(0)}=W_R+iW_I,
\end{equation}
where $W_R$ and $W_I$ are positive definite.
Then, Eq.~\eqref{eq_wkb_wo_grav} becomes
\begin{align}
    \frac{1}{2m}\sum_{a=-N}^N\qty[\qty(\pdv{W_R}{x^a})^2-\qty(\pdv{W_I}{x^a})^2]&=E-V(x^0),\\
    \sum_{a=-N}^N\qty(\pdv{W_R}{x^a})\qty(\pdv{W_I}{x^a})&=0.
\end{align}
The second equation shows that the degrees of freedom contributing to $W_R$ are orthogonal to those contributing to $W_I$.
Let us separate the first equation as 
\begin{align}
    \frac{1}{2m}\sum_{A}\qty(\pdv{W_R}{x_R^A})^2&=E-V(x^0)+\Xi,\label{eq_sep_xR}\\
    \frac{1}{2m}\sum_{B}\qty(\pdv{W_I}{x_I^B})^2&=\Xi,\label{eq_sep_xI}
\end{align}
where $\Xi$ is an arbitrary function and $x_R^A$ and $x_I^B$ are linear combinations of $x^a$. Here, $x_R^A$ contributes only to $W_R$ and $x_I^B$ contributes only to $W_I$. Since the left-hand side is positive definite, the equations may be integrated in the conventional way. However, $\Xi$ and the separation of $x_R^A$ and $x_I^B$ are not determined solely by the time-independent Schr\"odinger equation and can even depend on the integration parameter, $s$. This is a well-known problem in a quantum many-body system \cite{RAZAVY1988305} and occurs because the time-independent Schr\"odinger equation only has information about the total energy, whereas each particle has its own energy that cannot be transferred to another particle. 
We realize that this missing information is precisely the content of Eq.~\eqref{eq_sch_for_each}\footnote{Instead, one can determine them by constructing Huygens-type wave fronts sequentially from boundary conditions \cite{PhysRevA.41.32,doi:10.1063/1.466899,doi:10.1063/1.468526,doi:10.1063/1.3490087}. 
However, it is only possible along the solution of equations of motion \cite{PhysRevA.41.32} and cannot be extended to path integral formalism directly.}.

From the above observation, we consider a formulation based on the local energy conservation law, $\hat {\mathcal H}_a\psi=0$, rather than the time-independent Schr\"odinger equation, $(\hat H-E)\psi=0$.
This is also suggested by quantum gravity, where the corresponding constraints are called the Hamiltonian constraint and the momentum constraints.
Once we adopt this starting point, global time disappears from the theory since each particle can now choose Euclidean evolution or Lorentzian evolution independently. It would then not be possible to evolve a state using the time-dependent Schr\"odinger equation, which adopts global time. In the Heisenberg picture, this also means that we cannot rely on the Heisenberg equations of motion, $\dd{\hat A}/\dd{t}=i[\hat H,\hat A]/\hbar$,
where $\hat A$ denotes a generic operator. Thus, we do not follow the conventional procedure for deriving the path integral formula.

Finally, we note that quantum tunneling in a many-body system is often a result of interference and cancellation, which is difficult to capture in the conventional path integral and the WKB approximation. It is one of the reasons that the mixed tunneling is difficult to deal with.
We demonstrate this in Appendix \ref{apx_boundary_2d}.
\subsection{Path Integral for Separable System}
Following the above arguments, we redefine the problem as solving a system with constraints,
\begin{equation}
    \mathcal H_a\approx0,\label{eq_qm_const_sep}
\end{equation}
where $\approx$ indicates weak equality.

To treat Euclidean and Lorentzian evolution simultaneously, we introduce a time variable, $t^a$, for each particle. Then, we define the deformation parameter\footnote{It is also called the many-fingered or multi-fingered time in literature.}, $s$, as
\begin{equation}
    \dd{t^a}=\alpha^a(s)\dd{s}.\label{eq_sep_dta}
\end{equation}
Here, $s$ is real and $\alpha^a(s)$ is real or pure imaginary. Thus, $t^a$ is generally a complex variable. We will determine $\alpha^a(s)$ later and see that it becomes positive real for Lorentzian evolution and pure negative imaginary for Euclidean evolution.
The deformation parameter relates the time variables and allows us to define the equal-$s$ commutation relations as
\begin{equation}
    [\hat x^a(s),\hat p_b(s)]=i\hbar\delta^a_b.\label{eq_qm_xp}
\end{equation}
It generalizes the equal-time commutation relations to the operators with complex time variables.

Next, we find a way to evolve operators and derive a path integral formula.
We follow the Dirac's quantization approach for a constrained system; we first quantize the variables and solve the constraints by restricting Heisenberg states,
\begin{align}
    \hat {\mathcal H}_a(s)\ket{\psi}&=0.\label{eq_qm_lecl}
\end{align}
We first discretize $s$ and define $s_1,s_2,\dots,s_n$, where
\begin{equation}
    s_{J+1}=s_J+\delta s,
\end{equation}
with $\delta s>0$ being a small parameter. In the following, we use the notations like
\begin{equation}
    \hat x^a_J=\hat x^a(s_J),~\hat p_a^J=\hat p_a(s_J).
\end{equation}
These operators operate on the state as
\begin{align}
    \hat x^a_J\ket{\psi}&=\int\dd{X_J}\ket{X_J}\qty[x_J^a\bra{X_J}\ket{\psi}],\\
    \hat p_a^J\ket{\psi}&=\int\dd{X_J}\ket{X_J}\qty[\frac{\hbar}{i}\pdv{x_J^a}\bra{X_J}\ket{\psi}].
\end{align}
Here,
\begin{equation}
    \ket{X_J}=\ket{x_J^{-N},\dots,x_J^N},
\end{equation}
is an eigenstate of all $\hat x^a(s_J)$'s
and satisfies
\begin{equation}
    \int\dd{X_J}\ket{X_J}\bra{X_J}={\mathbb I},\label{eq_completeness}
\end{equation}
with ${\mathbb I}$ being identity.
We also defined
\begin{equation}
    \int\dd{X_J}=\int\prod_{a=-N}^N\dd{x^a_J}.
\end{equation}
Notice that $\ket{X_J}$ and $\ket{X_{J+1}}$ are different in general and we do not define the operations such as $\hat x^a_J\ket{X_{J+1}}$.
With these definitions, the $s$-evolution of the operators can be understood as that of the representation space.

Assuming $\braket{X_{J+1}}{X_J}$ is regular, we define $S_J$ as
\begin{align}
    \braket{X_{J+1}}{X_J}\bra{X_J}&=\exp{\frac{i}{\hbar}S_J(x_{J+1}^{-N}-x_J^{-N},\dots,x_{J+1}^N-x_J^N;X_{J+1})}\bra{X_J}.\label{eq_sep_def_sj}
\end{align}
Here, the last argument indicates the fixed parameters.
Notice that Eq.~\eqref{eq_completeness} implies $\ket{X_J}\braket{X_J}{X_{J+1}}=\ket{X_J}\exp(-iS_J/\hbar)$ and thus $\braket{X_J}{X_{J+1}}\neq\braket{X_{J+1}}{X_J}^*$ for a complex $S_J$.
From Eq.~\eqref{eq_qm_lecl}, we have
\begin{equation}
    \int\dd{X_{J+1}}\int\dd{X_J}\bra{\psi}\ket{X_{J+1}}\bra{X_{J+1}}\ket{X_J}\bra{X_J}\hat{\mathcal H}_a(s_J)=0.
\end{equation}
For this to be satisfied by arbitrary $\psi$, we require
\begin{equation}
    \ket{X_{J+1}}\bra{X_{J+1}}\ket{X_J}\bra{X_J}\hat{\mathcal H}_a(s_J)=0.\label{eq_sep_const_op}
\end{equation}
This gives the equation that involves $d^2S_J/(dx^a_J)^2$, but we are not interested in the second derivatives. Instead, we examine the trace\footnote{The trace of the equation has the following physical meaning. The path integral is formulated through the sequence of $\bra{\psi_f}\dots\ket{X_{J+1}}\bra{X_{J+1}}\ket{X_J}\bra{X_J}\ket{X_{J-1}}\bra{X_{J-1}}\dots\ket{\psi_i}$. Thus, what we really want to impose is
\begin{equation}
    \bra{X_{J+1}}\ket{X_J}\bra{X_J}\hat{\mathcal H_a}(s_J)\ket{X_{J-1}}=0.
\end{equation}
It requires the information of $X_{J+1},X_J$ and $X_{J-1}$ just like the second differential equation. Since we will take $X_{J+1}-X_J\to0$ and assume smooth paths ($C^1$-function almost everywhere), $S_J$ and $S_{J-1}$ becomes effectively the same. This justifies the evaluation of the trace.
} of Eq.~\eqref{eq_sep_const_op} and obtain
\begin{align}
    0&=\Tr[\ket{X_{J+1}}\bra{X_{J+1}}\ket{X_J}\bra{X_J}\hat{\mathcal H}_a(s_J)]\nonumber\\
    &=\int\dd{X'_J}\braket{X_{J+1}}{X_J}\bra{X_J}\qty[\frac{(\hat p_a^J)^2}{2m}+\delta_{a0}V(\hat x^a_J)-\mathcal E]\ket{X'_J}\bra{X'_J}\ket{X_{J+1}}\nonumber\\
    &=\braket{X_{J+1}}{X_J}\bra{X_J}\ket{X_{J+1}}\qty[\frac{1}{2m}\qty(\pdv{S_J}{x^a_J})^2+\delta_{a0}V(x^a_J)-\mathcal E].\label{eq_sep_dsdx_v}
\end{align}
We note that $\braket{X_{J+1}}{X_J}\bra{X_J}\ket{X_{J+1}}=1$.
Following Appendix \ref{apx_char}, we find that the characteristic curve is given by
\begin{equation}
    \frac{x^a_{J+1}-x^a_J}{\delta s}=-\frac{\alpha^a_J}{m}\pdv{S_J}{x^a_J},\label{eq_sep_dxds}
\end{equation}
where $\alpha^a_J=\alpha^a(s_J)$ is the Lagrange multiplier.
This gives the definition of $s$ and $\alpha^a$ in Eq.~\eqref{eq_sep_dta}.
Substituting Eq.~\eqref{eq_sep_dxds} into Eq.~\eqref{eq_sep_dsdx_v}, we obtain
\begin{equation}
    \pdv{S_J}{x^a_J}\frac{x^a_{J+1}-x^a_J}{\delta s}=-2\alpha_J^a\qty(\mathcal E-\delta_{a0}V(x^a_J)).
\end{equation}

From the Frobenius theorem, the system of these differential equations admits a solution of $S_J$ locally if, and only if, they have involutivity\footnote{If this were not satisfied, one would eventually get $S_J$ that is inconsistent with some of the constraints.}, {\it i.e.} $\mathcal L_i\mathcal L_j-\mathcal L_j\mathcal L_i=\sum_kc_{ij}^k\mathcal L_k$ for linear differential operators, $\mathcal L_i$.
With the prescription of
\begin{align}
    \qty(\hat p_a^J)^2&\to \frac{m}{\alpha_J^a}\frac{x^a_{J+1}-x^a_J}{\delta s}\frac{\hbar}{i}\pdv{x_J^a},\label{eq_sep_presc1}\\
    \hat p_a^J&\to \frac{1}{2}\qty(\frac{m}{\alpha_J^a}\frac{x^a_{J+1}-x^a_J}{\delta s}+\frac{\hbar}{i}\pdv{x_J^a}),\label{eq_sep_presc2}
\end{align}
it can be expressed as
\begin{equation}
    \braket{X_{J+1}}{X_J}\bra{X_J}[\hat{\mathcal H}_a(s_J),\hat{\mathcal H}_b(s_J)]\ket{X_{J+1}}=\sum_{c}\mathcal C_{ab}^c(X_J)\braket{X_{J+1}}{X_J}\bra{X_J}\hat{\mathcal H}_c(s_J)\ket{X_{J+1}},\label{eq_sep_involutivity}
\end{equation}
where $\mathcal C_{ab}^c(X_J)$'s are coefficients.
Notice that Eq.~\eqref{eq_sep_involutivity} is equivalent to
\begin{equation}
    \{\mathcal H_a,\mathcal H_b\}_P\approx0,
\end{equation}
where $\{*,*\}_P$ is the Poisson bracket. We call a subset of the constraints that satisfy the above equation as the involutive constraints. In this example, all the constraints are trivially involutive.

From Eqs.~\eqref{eq_sep_dsdx_v} and \eqref{eq_sep_dxds} , we also have
\begin{align}
    0&=\frac{m}{2(\alpha_J^a)^2}\qty(\frac{x^a_{J+1}-x^a_J}{\delta s})^2+\delta_{a0}V(x^a_J)-\mathcal E,
\end{align}
which can be solved\footnote{Even though $\mathcal H^a\approx0$ is difficult to solve in the space of $(x^a,p_a)$, it can be solved in $(x^a,dx^a/ds)$. This is because ``time'' is explicit in the Lagrangian formalism.} as
\begin{equation}
    (\alpha_J^a)^2=\frac{\frac{m}{2}\qty(\frac{x^a_{J+1}-x^a_J}{\delta s})^2}{\mathcal E-\delta_{a0}V(x^a_J)+i\epsilon}.
\end{equation}
Here, we chose an appropriate branch of $\alpha_J^a$ by adding an infinitesimal parameter, $\epsilon>0$. We will see that this corresponds to selecting the decaying solution during tunneling.
Notice that the degrees of freedom of the theory are not reduced by $\mathcal H^a\approx0$ since we solved it by the arbitrary Lagrange multiplier, $\alpha^a_J$.

Using these relations, we obtain
\begin{align}
    \braket{X_{J+1}}{X_J}&\simeq1-\frac{i}{\hbar}\sum_{a=-N}^N\pdv{S_J}{x^a_J}\frac{x^a_{J+1}-x^a_J}{\delta s}\delta s,\nonumber\\
    &=1+\frac{2i}{\hbar}\sum_{a=-N}^N\alpha_J^a(\mathcal E-\delta_{a0}V(x^a_J))\delta s\nonumber\\
    &=1+\frac{2i}{\hbar}\sum_{a=-N}^N\sqrt{\frac{m}{2}\qty(\frac{x_{J+1}^a-x_J^a}{\delta s})^2}\sqrt{\mathcal E-\delta_{a0}V(x^a_J)}\delta s.\label{eq_sep_xx}
\end{align}
Thus, we obtain the Green's function as
\begin{align}
    G(X_n,X_1;s_n-s_1)&=\braket{X_n}{X_1}\nonumber\\
    &=\int\qty(\prod_{J=2}^{n-1}\dd{X_J})\prod_{J=1}^{n-1}\qty[1+\frac{2i}{\hbar}\sum_{a=-N}^N\sqrt{\frac{m}{2}\qty(\frac{x_{J+1}^a-x_J^a}{\delta s})^2}\sqrt{\mathcal E-\delta_{a0}V(x^a_J)}\delta s]\nonumber\\
    &=\int\mathcal DX\exp\qty[\frac{i}{\hbar}\sum_{a=-N}^N\int_{s_1}^{s_n}\dd{s}2\sqrt{\frac{m}{2}\qty(\dv{x^a}{s})^2}\sqrt{\mathcal E-\delta_{a0}V(x^a)}],\label{eq_qm_sep_final}
\end{align}
where the infinite dimensional integral is denoted by $\int\mathcal DX$.
In the last line, we took $\delta s\to0$ and $n\to\infty$ with keeping $s_1$ and $s_n$ constant. More precisely, we also kept $(x^a_{J+1}-x^a_J)/\delta s$ regular as $\delta s\to0$, {\it i.e.} the paths are assumed to be continuous with respect to $s$.

We observe that the exponent is invariant under
\begin{equation}
    \dd{s}=J(s')\dd{s'},
\end{equation}
with $J(s')>0$ being an arbitrary function. This means that the path integral only remembers the order of the representation spaces and $(s_{J+1}-s_J)$ has no meaning. Thus, we rescale $s$ so that $s_1=0$ and $s_n=1$ without loss of generality.

The Green's function can be rewritten as
\begin{equation}
    G(X_n,X_1)=\int\mathcal DX\exp\qty[\frac{i}{\hbar}\sum_{a=-N}^N\int_{0}^{1}\dd{s}L^a],
\end{equation}
where
\begin{equation}
    L^a=\alpha^a \qty[\frac{m}{2}\qty(\frac{1}{\alpha^a}\dv{x^a}{s})^2-\delta_{a0}V(x^a)+\mathcal E].
\end{equation}
If $\alpha^a(s)=1$, our formula reproduces the Lorentzian path integral with $\dd{t}=\dd{s}$.
On the other hand, if $\alpha^a(s)=-i$, our formula reproduces the Euclidean path integral with $\dd{\tau}=\dd{s}$.
It is worth noting that our formula defines the path integral for a general $\alpha^a(s)$ and that the expression of Eq.~\eqref{eq_qm_sep_final} is valid even at the point where $\alpha^a$ becomes singular, which appears on the boundary between Euclidean and Lorentzian domains.

Let us define the transition probability using the Green's function.
The initial conditions at $s=0$ can be set through
\begin{equation}
    \braket{X_1}{\psi}=\int\dd{X_0}\braket{X_1}{X_0}\psi(X_0),
\end{equation}
where $\braket{X_1}{X_0}$ is defined through Eq.~\eqref{eq_sep_xx} although $\ket{X_0}$ is not necessarily defined.
Then, we obtain
\begin{equation}
    \braket{X}{\psi}=\int\dd{X_0}G(X,X_0)\psi(X_0),
\end{equation}
where $\ket{X}$ is the basis at $s=1$.
This is the wave function that the operators at $s=1$ observes. Thus, the differential probability to observe particles at $X$ at $s=1$ is given by
\begin{equation}
    dP(X)=\frac{\abs{\braket{X}{\psi}}^2}{\int\dd{X'}\abs{\braket{X'}{\psi}}^2}dX.
\end{equation}
Notice that the normalization of $\ket{\psi}$ is not conserved in general.

Now, we go back to the tunneling rate. 
We set the initial condition as
\begin{equation}
    \psi(x^{-N},\dots,x^N)=\prod_{a=-N}^N\delta(x^a-x_{\rm ini}),
\end{equation}
and observe the particles at
\begin{equation}
    X=(x_{\rm fin},\dots,x_{\rm fin}),
\end{equation}
where $x_{\rm ini}<0$ and $x_{\rm fin}>1$.
The dominant contribution of the path integral comes from a path with constant $dx^a/ds$ because it apparently minimizes both the real part and the imaginary part of the exponent. Then, we can see that the exponent agrees with that of Eq.~\eqref{eq_correct_rate}.

Before closing this section, let us see what happens if we impose only $(\hat H(s)-E)\ket{\psi}=0$ and follow the same procedure. We immediately find that we cannot determine all of $\alpha^a$'s only with this constraint. If we assume the existence of global time, we can make $\alpha^a$ independent of $a$ as explained in Appendix \ref{apx_char}. Then, we obtain the conventional Lorentzian or Euclidean path integral formula depending on the sign of $(V(x^0)-E)$. For $V_0<E$, the tunneling probability appears to be one.
This reminds us the wrong result we saw in Subsection \ref{subsec_lecl}, where we solved $(\hat H-E)\psi=0$ in the WKB approximation. This implies that we need to determine all $\alpha^a$'s properly to see the mixed tunneling explicitly and it requires the same number of equations. In our formulation, they are provided by the local energy conservation law. Notice that this does not mean that the conventional Lorentzian path integral cannot describe the mixed tunneling. As we will explain in Section \ref{sec_relation}, our path integral formula can be understood from the conventional one by considering interference effects.

\section{Non-separable System}\label{sec_non_separable}
\setcounter{equation}{0}
Let us move on to a non-trivial problem, where there are interactions between the neighbors. The Hamiltonian is extended as
\begin{equation}
    H=\sum_{a=-N}^N\frac{p_a^2}{2m}+V(x^0)+\sum_{a=-N}^{N-1}\frac{m\omega^2}{2}(x^{a+1}-x^a)^2,
\end{equation}
where $\omega$ is a constant. Here, we have chosen specific interactions so that the discussion becomes parallel to that for quantum field theory.
For another example, see Appendix~\ref{apx_two_body}.
\subsection{Local Energy Conservation Law}
We first find a local energy conservation law of this system.
The Hamiltonian dynamics determines the time evolution of $x^a$ and $p_a$ as
\begin{align}
    \dv{x^a}{t}&=\frac{p_a}{m},\label{eq_evol_x}\\
    \dv{p_a}{t}&=-\delta_{a0}V'(x^a(t))-m\omega^2(x^a(t)-x^{a+1}(t))-m\omega^2(x^a(t)-x^{a-1}(t)),\label{eq_evol_p}
\end{align}
for $-N<a<N$.
We omit the equations at the edges here and after.
Substituting these into the right-hand side of
\begin{align}
    \frac{1}{2m}\qty[p_a^2(t_f)-p_a^2(t_i)]&=\int_{t_i}^{t_f}\dd{t}\frac{p_a(t)}{m}\dv{p_a(t)}{t},
\end{align}
we obtain
\begin{align}
    0&=\frac{p_a^2(t)}{2m}+\delta_{a0}V(x^a(t))+\frac{m\omega^2}{2}\frac{(x^a(t)-x^{a+1}(t))^2+(x^a(t)-x^{a-1}(t))^2}{2}\nonumber\\
    &\hspace{3ex}-\frac{m\omega^2}{2}\int_0^t\dd{t'}\qty(\dv{x^{a+1}(t')}{t}+\dv{x^a(t')}{t})(x^{a+1}(t')-x^a(t'))\nonumber\\
    &\hspace{3ex}+\frac{m\omega^2}{2}\int_0^t\dd{t'}\qty(\dv{x^a(t')}{t}+\dv{x^{a-1}(t')}{t})(x^a(t')-x^{a-1}(t'))-\mathcal E^a,\label{eq_qm_local_energy_cons}
\end{align}
where $\mathcal E^a$ is a constant.
This represents the local energy conservation law as we see below. The first line is the local kinetic energy, potential energy, and the half of the energy of the nearby springs. The second line can be understood as the energy increase of the $a$-particle by the $(a,a+1)$-spring (minus the energy increase of the half spring).
Thus, the last two lines describe time-dependent local energy (but not including that of the springs), which accumulates all the energy transferred into the $a$-particle.

Next, we rewrite the integrals in Eq.~\eqref{eq_qm_local_energy_cons} in a more convenient form. We introduce two sets of auxiliary operators, $\eta_{\pm}^a(s)$, and their momentum conjugates, $\pi^{\pm}_a(s)$.
Then, we consider the following constraints;
\begin{equation}
    \mathcal H_a\approx0,~\mathcal H_a^\pm\approx0,
\end{equation}
where
\begin{align}
    \mathcal H_a&=\frac{p_a^2}{2m}+\frac{1}{2}{(\pi^+_a)^2}+\frac{1}{2}{(\pi^-_a)^2}+\delta_{a0}V(x^a)\nonumber\\
    &\hspace{3ex}+\frac{m\omega^2}{2}\frac{(x^a-x^{a+1})^2+(x^a-x^{a-1})^2}{2}-\frac{\eta^{a+1}_-+\eta^a_+}{2\kappa}+\frac{\eta^a_-+\eta^{a-1}_+}{2\kappa},\label{eq_nonsep_const1}\\
    \mathcal H_a^+&=\omega^2(x^{a+1}-x^a)p_a-\frac{1}{\kappa}\pi^{+}_a,\label{eq_nonsep_const2}\\
    \mathcal H_a^-&=\omega^2(x^a-x^{a-1})p_a-\frac{1}{\kappa}\pi^{-}_a.\label{eq_nonsep_const3}
\end{align}
Here, $\kappa$ is a constant having mass dimension $-3/2$. We introduced kinetic terms for $\eta^a_\pm$ so that $\{\mathcal H_a,\mathcal H_b\}_P\approx0$ is satisfied, and two auxiliary operators instead of one to ensure that $x^a$ is always a real number.
We absorbed $\mathcal E^a$ in the initial conditions of $\eta_\pm^a$.
We will take the limit of $\kappa\to0$ at the end to get rid of the effect of the kinetic term of $\eta^a_\pm$\footnote{Alternatively, one can introduce another set of variables to eliminate the degrees of freedom of $\eta^a_\pm$ as in Appendix \ref{apx_two_body}.
The structure of the kinetic term of $\eta^a_\pm$ is undetermined unless we include gravity. Since any assumption on the kinetic term gives the same result after $\kappa\to0$, we proceed with the minimal setup.}.
In this example, we have $\{\mathcal H_a,\mathcal H_b^\pm\}_P\not\approx0$ and $\{\mathcal H_a^\pm,\mathcal H_b^\pm\}_P\not\approx0$, and thus only $\mathcal H_a\approx0$ is the involutive constraint.

Before closing this subsection, we comment on the choice of the energy conservation equations, Eq.~\eqref{eq_qm_local_energy_cons}.
It corresponds to finding what are the right degrees of freedom that store the local energy.
For example, in a system with two degrees of freedom, $x^0$ and $x^1$, we may derive two conservation equations for $x^0$ and $x^1$, or for $(x^0+x^1)$ and $(x^0-x^1)$. It is also possible to derive a single conservation equation for $x^0$ and $x^1$.
If we do not consider gravity, we may choose\footnote{We do not have a suggestion of the right choice to be confident about for generic theories other than quantum gravity and quantum field theory.} a convenient set of equations that is convenient for the situation as long as there are a sufficient number of involutive constraints.
However, once we include gravity, the curvature of space is determined by the energy at each spatial point. Thus, energy has to be well-defined at each spatial point. In fact, the energy conservation equations cannot be freely chosen in quantum gravity, and the corresponding constraints are given at each spatial point by the definition of the theory.
If we assume quantum field theory is obtained in the decoupling limit of gravity, this should be also true for quantum field theory.

\subsection{Path Integral for Non-separable System}
We follow the formalism in the separable example and obtain the path integral formula for the non-separable system. In this and the subsequent sections, we use the same variables as those in the previous section in order to avoid cumbersome notation.

We first introduce the deformation variable, $s$, and set the commutation relations as
\begin{equation}
    [\hat x^a(s),\hat p_b(s)]=[\hat \eta_\pm^a(s),\hat \pi^\pm_b(s)]=i\hbar\delta^a_b.
\end{equation}

Next, we discretize $s$ and define a basis at $s=s_J$ as
\begin{equation}
    \ket{X_J}=\ket{x_J^{-N},\dots,x_J^N,(\eta^{-N}_{\pm})_J,\dots,(\eta^N_{\pm})_J}.
\end{equation}
Here, $(\hat\eta_{\pm}^a)_J$ and $(\hat\pi^{\pm}_a)^J$ operate on the state in the similar way as $\hat x^a_J$ and $\hat p_a^J$.
Then, we define $S_J$ through
\begin{equation}
    \braket{X_{J+1}}{X_J}\bra{X_J}=\exp{\frac{i}{\hbar}S_J(X_{J+1}-X_J;X_{J+1})}\bra{X_J}.
\end{equation}

From the constraint of Eq.~\eqref{eq_nonsep_const1}, we have
\begin{align}
    0&=\frac{1}{2m}\qty(\pdv{S_J}{x^a_J})^2+\frac12\qty(\pdv{S_J}{\eta_{+J}^a})^2+\frac12\qty(\pdv{S_J}{\eta_{-J}^a})^2+\delta_{a0}V(x^a_J)\nonumber\\
    &\hspace{3ex}+\frac{m\omega^2}{2}\frac{( x^a_J- x^{a+1}_J)^2+(x^a_J- x^{a-1}_J)^2}{2}-\frac{(\eta^{a+1}_{-})_J+(\eta^a_{+})_J}{2\kappa}+\frac{(\eta^a_{-})_J+(\eta^{a-1}_{+})_J}{2\kappa}.
\end{align}
One of the differences from the separable example is that there is more than one variable for each $a$.
As explained in Appendix \ref{apx_char}, we can in principle introduce different time variables for $x^a$ and $\eta^a_\pm$ to make the solution as general as possible:
\begin{align}
    \delta t_x&=\frac{x^a_{J+1}-x^a_J}{-\frac{1}{m}\pdv{S_J}{x^a_J}},\\
    \delta t_{\eta_\pm}&=\frac{(\eta^a_{\pm})_{J+1}-(\eta^a_{\pm})_J}{-\pdv{S_J}{\eta_{\pm J}^a}}.
\end{align}
Notice that, according to this definition, time is measured by how far a variable has traveled for a given velocity. Hence, it is a proper time for the variable.
Then, the constraint of Eq.~\eqref{eq_nonsep_const2} becomes
\begin{align}
    0&=\Tr[\ket{X_{J+1}}\bra{X_{J+1}}\ket{X_J}\bra{X_J}\hat{\mathcal H}_a^+(s_J)]\nonumber\\
    &=\int\dd{X'_J}\bra{X_{J+1}}\ket{X_J}\bra{X_J}\qty[\omega^2(\hat x^{a+1}_J-\hat x^a_J)\hat p_a^J-\frac{1}{\kappa}\hat\pi^{+J}_a]\ket{X'_J}\bra{X'_J}\ket{X_{J+1}}\nonumber\\
    &= m\omega^2(x^{a+1}_J-x^a_J)\frac{x^a_{J+1}-x^a_J}{\delta t_x}-\frac{1}{\kappa}\frac{(\eta^a_+)_{J+1}-(\eta^a_+)_J}{\delta t_{\eta+}}.
\end{align}
Thus, we have
\begin{equation}
    \frac{1}{\kappa}\qty[(\eta^a_+)_{J+1}-(\eta^a_+)_J]=m\omega^2(x^{a+1}_J-x^a_J)(x^a_{J+1}-x^a_J)\frac{\delta t_{\eta+}}{\delta t_x}.\label{eq_non_sep_enegy_dt}
\end{equation}
By construction, the left-hand side can be understood as a change in the local energy. The right-hand side is essentially the force acting on $x^a$ multiplied by the distance that traveled by $x^a$, {\it i.e.} the work done by $x^a$. However, for this to be the local energy conservation law, we need $\delta t_x=\delta t_{\eta_+}$.
This equality is also important because, if each variable had its own time, we could obtain a different local energy for a different choice of ``equal-time slice'', and the local energy itself would become meaningless. More technically speaking, if $\delta t_x$ were pure imaginary and $\delta t_{\eta_+}$ were real, Eq.~\eqref{eq_non_sep_enegy_dt} becomes inconsistent. The same is also true for $\eta^a_-$. From these reasons, we introduce only one time variable for each $a$ as part of the definition of the energy conservation law.
Then, we define
\begin{equation}
    \delta t_x=\delta t_{\eta_\pm}=\alpha_J^a\delta s.
\end{equation}
This means that we can use the method of characteristics safely.
The characteristic curve is given by
\begin{align}
    \frac{x^a_{J+1}-x^a_J}{\delta s}&=-\frac{\alpha_J^a}{m}\pdv{S_J}{x^a_J},\\
    \frac{(\eta^a_\pm)_{J+1}-(\eta^a_\pm)_J}{\delta s}&=-\alpha_J^a\pdv{S_J}{\eta_{\pm J}^a}.
\end{align}

Now, let us go back to Eq.~\eqref{eq_non_sep_enegy_dt} with $\delta t_x=\delta t_{\eta_+}$. It restricts $((\eta^a_{+})_{J+1}-(\eta^a_{+})_J)$ for a given $(x^a_{J+1}-x^a_J)$.
Together with the constraint of Eq.~\eqref{eq_nonsep_const3}, the trajectories of $\eta_{\pm}^a(s)$ are determined uniquely and thus the degrees of freedom of $\eta_{\pm}^a$ are eliminated.
In the following, we only consider the paths that satisfy these constraints by constraining the representation space. Then, $S_J$ becomes a function of $(x^a_{J+1}-x^a_J)$,
\begin{equation}
    S_J(X_{J+1}-X_J;X_{J+1})=S_J(x_{J+1}^{-N}-x_J^{-N},\dots,x_{J+1}^N-x_J^N;X_{J+1}),
\end{equation}
and $\hat{\mathcal H}^\pm_a$ become the directional derivatives of $S_J$ perpendicular to the path. Since there is no representation space off the path, we can set $\hat{\mathcal H}^\pm_a=0$ at the operator level\footnote{By eliminating $\eta_\pm^a$, we are essentially going back to the original constraint of Eq.~\eqref{eq_qm_local_energy_cons}, and there is no $\hat{\mathcal H}^\pm_a$ from the beginning.}.
Thus, $(\hat\pi^{\pm}_a)^J$ are identified as
\begin{align}
    (\hat\pi^{+}_a)^J&=\kappa\omega^2(\hat x^{a+1}_J-\hat x^a_J)\hat p_a^J,\\
    (\hat\pi^{-}_a)^J&=\kappa\omega^2(\hat x^a_J-\hat x^{a-1}_J)\hat p_a^J,
\end{align}
on the constrained representation space. 

As in the previous example, the constraint of Eq.~\eqref{eq_nonsep_const1} can be solved by adjusting $\alpha^a_J$ as
\begin{equation}
    (\alpha^a_J)^2=\frac{\mathcal K^a_J}{-\mathcal V^a_J+i\epsilon},
\end{equation}
where
\begin{align}
    \mathcal K^a_J&=\frac{m}{2}\qty(\frac{x^a_{J+1}-x^a_J}{\delta s})^2+\order{\kappa^2},\\
    \mathcal V^a_J&=\delta_{a0}V(x^a_J)+\frac{m\omega^2}{2}\frac{(x^a_J-x^{a+1}_J)^2+(x^a_J-x^{a-1}_J)^2}{2}\nonumber\\
    &\hspace{3ex}-\frac{(\eta^{a+1}_-)_J+(\eta^a_+)_J}{2\kappa}+\frac{(\eta^a_-)_J+(\eta^{a-1}_+)_J}{2\kappa}.
\end{align}
Since $\mathcal H_a$'s are involutive, we can integrate $S_J$ consistently and obtain
\begin{align}
    \braket{X_{J+1}}{X_J}&\simeq1-\frac{i}{\hbar}\sum_{a=-N}^N\pdv{S_J}{x^a_J}\frac{x^a_{J+1}-x^a_J}{\delta s}\delta s\nonumber\\
    &\simeq1+\frac{2i}{\hbar}\sum_{a=-N}^N\sqrt{\mathcal K^a_J}\sqrt{-\mathcal V^a_J}\delta s.
\end{align}
Thus, the Green's function is obtained as
\begin{align}
    G(X_n,X_1)&=\int\qty(\prod_{J=2}^{n-1}\dd{X_J}\delta_{\eta_J^+}\delta_{\eta_J}^-)\prod_{J=1}^{n-1}\qty[1+\frac{2i}{\hbar}\sum_{a=-N}^N\sqrt{\mathcal K^a_J}\sqrt{-\mathcal V^a_J}\delta s]\nonumber\\
    &=\int\mathcal DX\delta_\eta\exp\qty[\frac{i}{\hbar}\sum_{a=-N}^N\int_{0}^{1}\dd{s}2\sqrt{\mathcal K^a(s)}\sqrt{-\mathcal V^a(s)}],\label{eq_qm_ns_path_int}
\end{align}
where
\begin{equation}
    \delta_{\eta_J^+}=\prod_{a=-N}^{N-1}\delta\qty((\eta^a_+)_{J+1}-(\eta^a_+)_J-\kappa m\omega^2(x^{a+1}_J-x^a_J)(x^a_{J+1}-x^a_J)),
\end{equation}
and the similar definition for $\delta_{\eta_J^-}$. 
In the second line, we combined $\eta_{\pm}^a$ as
\begin{equation}
    \eta_J^a=\frac12\qty[(\eta_+^a)_J+(\eta_-^{a+1})_J],
\end{equation}
and took $\delta s\to0$.
The constrained path integral, $\int\mathcal DX\delta_\eta$, denotes the integration over the paths satisfying
\begin{align}
    \dv{\eta^a(s)}{s}&=\frac{\kappa m\omega^2}{2}(x^{a+1}(s)-x^a(s))\qty(\dv{x^{a+1}(s)}{s}+\dv{x^a(s)}{s}).
\end{align}
The integrand in the second line is given by
\begin{align}
    \mathcal K^a&=\frac{m}{2}\qty(\dv{x^a}{s})^2,\\
    \mathcal V^a&=\delta_{a0}V(x^a)+\frac{m\omega^2}{2}\frac{(x^a-x^{a+1})^2+(x^a-x^{a-1})^2}{2}-\frac{\eta^a-\eta^{a-1}}{\kappa}.
\end{align}

For the special cases where $\alpha^a(s)=1$ ($\alpha^a(s)=-i$), Eq.~\eqref{eq_qm_ns_path_int} becomes the conventional Lorentzian (Euclidean) path integral.

\section{Quantum Field Theory}\label{sec_qft}
\setcounter{equation}{0}
It is straightforward to extend the previous argument to quantum field theory. We consider the Hamiltonian given by
\begin{equation}
    H=\int\dd[3]{x}\qty[\frac12\pi^2(\bfx)+\frac12(\partial_i\phi(\bfx))(\partial^i\phi(\bfx))+V(\phi(\bfx))],\label{eq_qft_orig_h}
\end{equation}
where $\phi$ is a scalar field and $\pi$ is its momentum conjugate.
Here after, the summation over the spatial index is implicit and we raise and lower it with $\delta_{ij}$.

In the following, we formulate path integral and then obtain the equations of motion with the constraints. Then, we also analyze a system with more involtive constraints.
\subsection{Path Integral for Quantum Field Theory}\label{subsec_qft}
The local energy conservation law can be found in the same way as in the previous example and then the system is defined through constraints,
\begin{equation}
    \mathcal H\approx0,~{\mathcal H}_i\approx0,
\end{equation}
where
\begin{align}
    \mathcal H&=\frac12\pi^2+\frac12\pi^\eta_i\pi^{\eta i}+\frac12(\partial_i\phi)(\partial^i\phi)+V(\phi)-\frac{1}{\kappa}\partial_i\eta^i,\label{eq_qft_ham_const}\\
    \mathcal H_i&=(\partial_i\phi)\pi-\frac{1}{\kappa}\pi^\eta_i.\label{eq_qft_mom_const}
\end{align}
Here, $\kappa$ is a constant having mass dimension $-2$. We introduced three auxiliary fields, $\eta^i(\bfx)$'s, and their conjugates, $\pi^\eta_i(\bfx)$'s.
In quantum field theory, there are several options for the auxiliary fields. For example, we could use\footnote{We need at least three degrees of freedom to store all the information. Thus, it is not possible to use only $\partial^i\eta$ unless a spatial symmetry is assumed.} $\partial^i\eta$ or $\partial_j\eta^{ij}$ instead of $\eta^i$. A three vector, $\eta^i$, is the simplest option and facilitates the restoration of Lorentz covariance. On the other hand, the structure deduced from quantum gravity is $\qty(\partial_j\eta^{ij}-\partial^i\eta_j^j)$, which involves more degrees of freedom and requires a gravitational Hamiltonian to restore Lorentz covariance. Each of them is self-consistent, but gives slightly different results. We will go back to this point in Subsection \ref{subsection_decoupling}.

The poisson brackets are defined through the smeared operators like
\begin{equation}
    \mathcal H_i[g^i]=\int\dd[3]{x}g^i(\bfx)\mathcal H_i(\bfx).
\end{equation}
We find $\{\mathcal H[f],\mathcal H[g]\}_P\approx0$, $\{\mathcal H[f],\mathcal H_i[g^i]\}_P\not\approx0$ and $\{\mathcal H_i[f^i],\mathcal H_j[g^j]\}_P\not\approx0$. Thus, only $\mathcal H$ has involutivity.

We set the equal-$s$ commutation relations as
\begin{equation}
    [\hat\phi(s,\bfx),\hat\pi(s,\bfy)]=\delta^3(\bfx-\bfy),~[\hat\eta^i(s,\bfx),\hat\pi^{\eta}_j(s,\bfy)]=\delta^i_j\delta^3(\bfx-\bfy).
\end{equation}
In the following, we avoid defining similar quantities that have appeared in the previous sections.

We integrate $S_J$ along the characteristic curve, which is given by
\begin{align}
    \frac{\phi_{J+1}(\bfx)-\phi_J(\bfx)}{\delta s}&=-\alpha_J(\bfx)\pdv{S_J}{\phi_J(\bfx)},\\
    \frac{\eta^i_{J+1}(\bfx)-\eta^i_J(\bfx)}{\delta s}&=-\alpha_J(\bfx)\pdv{S_J}{\eta_J^i(\bfx)},
\end{align}
with $\alpha_J(\bfx)$ being the Lagrange multiplier. After taking $\delta s\to0$ and $\kappa\to0$, we obtain
\begin{equation}
    \alpha^2(s,\bfx)=\frac{\mathcal K(s,\bfx)}{-\mathcal V(s,\bfx)+i\epsilon},
\end{equation}
where
\begin{align}
    \mathcal K(s,\bfx)&=\frac12\qty(\partial_s\phi(s,\bfx))^2,\label{eq_qft_k}\\
    \mathcal V(s,\bfx)&=V(\phi(s,\bfx))+\frac12(\partial_i\phi(s,\bfx))(\partial^i\phi(s,\bfx))-\frac{1}{\kappa}\partial_i\eta^i(s,\bfx).\label{eq_qft_v}
\end{align}

The Green's function is derived as
\begin{align}
    G(\phi_n,\eta_n,\phi_1,\eta_1)&=\int\mathcal D\phi\mathcal D\eta\delta_\eta\exp\qty[\frac{i}{\hbar}\int_{0}^{1}\dd{s}\int\dd[3]{x}2\sqrt{\mathcal K(s,\bfx)}\sqrt{-\mathcal V(s,\bfx)}],\label{eq_qft_final}
\end{align}
where the path integral is executed over the paths satisfying
\begin{equation}
    \partial_s\eta^i(s,\bfx)=\kappa(\partial^i\phi(s,\bfx))(\partial_s\phi(s,\bfx)).\label{eq_qft_dsdeta}
\end{equation}
We note that this result is for a particular choice of the auxiliary fields and the result deduced from quantum gravity is given in Subsection \ref{subsection_decoupling}.

Let us define a four-dimensional metric,
\begin{equation}
    g_{\mu\nu}\dd{x^\mu}\dd{x^\nu}=-\alpha^2\dd{s}^2+\dd{x^i}\dd{x_i},
\end{equation}
and the $s$-component of $\eta^\mu$,
\begin{equation}
    g^{ss}\partial_s\eta_s=\kappa g^{ss}(\partial_s\phi)^2=-\frac{\kappa}{\alpha^2}(\partial_s\phi)^2.
\end{equation}
Then, the exponent of the path integral can be written in the Lorentz invariant form as
\begin{align}
    \mathcal L=2\sqrt{\mathcal K}\sqrt{-\mathcal V}=\sqrt{-g}\qty[-\frac12g^{\mu\nu}(\partial_\mu\phi)(\partial_\nu\phi)-V(\phi)+\frac{1}{\kappa}\qty(u^\mu u^\nu-u^\rho u_\rho g^{\mu\nu})\partial_\mu \eta_\nu],\label{eq_qft_lagrangian}
\end{align}
where
\begin{equation}
    u_\mu=(\alpha,0,0,0).\label{eq_def_u}
\end{equation}
If $\alpha^2(s,\bfx)=\pm1$, the path integral is identical to Euclidean or Lorentzian path integral except for the last term, which gives the total energy of the system after integrating over space. Notice that $u_\mu$ remembers the Lorentz frame where the initial local energy, {\it i.e.} the initial conditions of $\eta^i$, is set. We could do it in a different Lorentz frame as the Lorentz transformation ensures a space-like point remains space-like. Then, $\tilde\eta^\mu$ with a different $\tilde u^\mu$ is related to the original variables through
\begin{equation}
    \qty(\tilde u^\mu \tilde u^\nu-\tilde u^\rho \tilde u_\rho g^{\mu\nu})\partial_\mu \tilde \eta_\nu=\qty(u^\mu u^\nu-u^\rho u_\rho g^{\mu\nu})\partial_\mu \eta_\nu.
\end{equation}

Notice that it is not a valid question whether the sign of $\alpha^2$ is Lorentz invariant. The Lorentz symmetry is an isometry, and $\alpha^2$ is the $(ss)$-component of the metric. Thus, it is preserved by definition. A Euclidean domain seen from Lorentzian spacetime is a defect of spacetime. 
\subsection{Equations of Motion}
Let us examine the equations of motion under the constraints of Eq.~\eqref{eq_qft_dsdeta}. Here, the difficulty is that the constraints are not Lorentz covariant. The following describes how to recover the covariance and obtain the conventional equations of motion.

We first find a symmetry of the theory.
Let us consider the transformation defined by
\begin{align}
    g_{\mu\nu}\to \tilde g_{\mu\nu}(\boldsymbol\lambda)=(\Lambda^{-1})_\mu^{~\rho}(\Lambda^{-1})_\nu^{~\sigma}g_{\rho\sigma},\label{eq_qft_metric_transform}
\end{align}
where
\begin{equation}
    \Lambda^{~s}_s=1,~\Lambda^{~s}_i=0,~\Lambda^{~i}_s=-\lambda_i,~\Lambda^{~i}_j=\delta^i_j.
\end{equation}
Notice that this transformation is equivalent to choosing a different four-dimensional metric when we rewrite Eq.~\eqref{eq_qft_final} to the four-dimensional form.

Let $\mathcal L$ with $\tilde g_{\mu\nu}(\boldsymbol\lambda)$ be denoted by $\mathcal L_{\boldsymbol\lambda}$.
Since $u_\mu$ and $\eta_i$ are unchanged by the multiplication of $\Lambda^{-1}$, $\mathcal L_{\boldsymbol\lambda}$ is identical to $\mathcal L$ after the coordinate transformation,
\begin{equation}
    \partial_\mu\to(\Lambda^{-1})^{~\nu}_\mu\partial_\nu.
\end{equation}
Since the Jacobian of the transformation is one, we have
\begin{equation}
    \int\dd[4]{x}\mathcal L=\int\dd[4]{x}\mathcal L_{\boldsymbol\lambda},
\end{equation}
ignoring the boundary, which is irrelevant in deriving the equations of motion.

To extremize the exponent of Eq.~\eqref{eq_qft_final} with the constraints, 
we introduce Lagrange multipliers, $\beta^i(s,\bfx)$, and define
\begin{align}
    S_{\boldsymbol\beta}&=\int\dd[4]{x}\qty[\mathcal L-\frac{2\beta^i}{\alpha}\qty((\partial_i\phi)(\partial_s\phi)-\frac{\partial_s\eta_i}{\kappa})]\nonumber\\
    &=\int\dd[4]{x}\qty[\mathcal L_{-\boldsymbol\beta}-\frac{2\beta^i}{\alpha}\qty((\partial_i\phi)(\partial_s\phi)-\frac{\partial_s\eta_i}{\kappa})]\nonumber\\
    &=\int\dd[4]{x}\mathcal L_{\boldsymbol\beta}.
\end{align}

By variating $S_{\boldsymbol\beta}$ with respect to $\beta^i$, we obtain
\begin{align}
    \delta S_{\boldsymbol\beta}&=\frac{1}{\alpha^2}\qty{\beta^j\qty[(\partial_i\phi)(\partial_j\phi)-\frac{1}{\kappa}(\partial_i\eta_j+\partial_j\eta_i)]-\qty[(\partial_i\phi)(\partial_s\phi)-\frac{1}{\kappa}\partial_s\eta_i]}\delta\beta^i,\label{eq_qft_eom_beta}
\end{align}
which determines the Lagrange multiplier, $\beta^j$. Here, the terms proportional to $dS_{\boldsymbol\beta}/d\alpha$ vanished due to the constraints.

We skip the equations of motion for $\eta^i$ since we have omitted $\order{\kappa^2}$ terms in the Lagrangian.
By variating $S_{\boldsymbol\beta}$ with respect to $\phi$, we obtain
\begin{align}
    \delta S_{\boldsymbol\beta}&=\sqrt{-\tilde g}\qty[-\tilde D_\mu\partial^\mu\phi-V'(\phi)]\delta\phi,
\end{align}
where $\tilde D_\mu$ is the covariant derivative for $\tilde g_{\mu\nu}$.
Thus, with $\alpha^2=\pm1$, this agrees with the conventional Euclidean or Lorentzian equations of motion.

\subsection{System with More Involutive Constraints}\label{subsec_more_ic}
So far, we have analyzed only systems that have the same number of involutive constraints as that of the physical degrees of freedom.
It is thus instructive to see a system with more involutive constraints.
We extend the previous system so that $\{\mathcal H,\mathcal H_i\}\approx0$ and find that
the minimally extended system is defined through the following constraints;
\begin{equation}
    \mathcal H\approx0,~{\mathcal H}_i\approx0,~\pi^\zeta\approx0,\label{eq_qft_constraints}
\end{equation}
where
\begin{align}
    \mathcal H&=\frac12\pi^2+\frac12\pi^\eta_i\pi^{\eta i}+\frac12(\partial_i\phi)(\partial^i\phi)+\frac12(\partial_i\eta^j)(\partial^i\eta_j)+V(\phi)-\frac{1}{\kappa}\partial_i\eta^i-\zeta+\frac{1}{\kappa^2},\\
    \mathcal H_i&=(\partial_i\phi)\pi+(\partial_i\eta^j)\pi^\eta_j+(\partial_i\zeta)\pi^\zeta-\frac{1}{\kappa}\pi^\eta_i-\partial_j\qty(T^j_i\pi^\zeta),
\end{align}
with
\begin{align}
    T^k_l&=\delta^k_l\qty[\frac12\pi^2+\frac12\pi^\eta_i\pi^{\eta i}-\frac12(\partial_i\phi)(\partial^i\phi)-\frac12(\partial_i\eta^j)(\partial^i\eta_j)-V(\phi)+\frac{1}{\kappa}\partial_i\eta^i+\zeta]\nonumber\\
    &\hspace{3ex}+(\partial_l\phi)(\partial^k\phi)+(\partial_l\eta^i)(\partial^k\eta_i)-\frac{1}{\kappa}\partial_l\eta^k-\frac{1}{\kappa}\partial^k\eta_l.
\end{align}
Here, we added an auxiliary field, $\zeta$, and its momentum conjugate, $\pi^\zeta$.
We note that this is not necessarily a physically meaningful model.
The Poisson brackets of the constraints are given by\footnote{The constraint algebra of the first three is identical to that in pure gravity \cite{diraclecture}.}
\begin{align}
    \{\mathcal H[f],\mathcal H[g]\}_P&\stackrel{\pi^\zeta}{\approx}\mathcal H_i[f\partial^ig-g\partial^if]\approx0,\\
    \{\mathcal H[f],\mathcal H_i[g^i]\}_P&\stackrel{\pi^\zeta}{\approx}\mathcal H[-g^i\partial_if]\approx0,\\
    \{\mathcal H_i[f^i],\mathcal H_j[g^j]\}_P&\stackrel{\pi^\zeta}{\approx}\mathcal H_i[f^j\partial_jg^i-g^j\partial_jf^i]\approx0,\\
    \{\mathcal H[f],\pi^\zeta[g]\}_P&\approx I[-fg],
\end{align}
where $\stackrel{\pi^\zeta}{\approx}$ is the equality after $\pi^\zeta\approx0$ is imposed and 
\begin{equation}
    I[f]=\int\dd[3]{x}f(\bfx).
\end{equation}
Thus, both $\mathcal H$ and $\mathcal H_i$'s become involutive operators.

Let us obtain the path integral formula.
In addition to $\hat\phi$ and $\hat\eta^i$, we set the equal-$s$ commutation relations for $\hat \zeta$ as
\begin{equation}
    [\hat\zeta(s,\bfx),\hat\pi^\zeta(s,\bfy)]=\delta^3(\bfx-\bfy).
\end{equation}
Then, we discretize $s$ and define a basis at $s=s_J$ as
\begin{equation}
    \ket{X_J}=\ket{\phi_J,\eta_J,\zeta_J}.
\end{equation}
Here after, we omit the index of $\eta^i$ in the state.

Since $\pi^\zeta\approx0$ simply eliminates $\hat\zeta$, we have
\begin{align}
    \zeta_{J+1}-\zeta_J=0,~\hat\pi^{\zeta J}=0.
\end{align}
Thus, we define $S_J$ as
\begin{equation}
    \braket{X_{J+1}}{X_J}\bra{X_J}=\exp{\frac{i}{\hbar}S_J(\phi_{J+1}-\phi_J,\eta_{J+1}-\eta_J;X_{J+1})}\bra{X_J}.
\end{equation}
As a constant shift of $\zeta$ can be absorbed by a redefinition of $\eta^i$, we take $\zeta=1/\kappa^2$ without loss of generality.

Since both $\mathcal H$ and $\mathcal H_i$'s have involutivity, we can consistently integrate $S_J$ without constraining $\eta^i$'s. As for the differential equations for $S_J$, we have
\begin{align}
    0&=\Tr\qty[\ket{X_{J+1}}\braket{X_{J+1}}{X_J}\bra{X_J}\hat{\mathcal H}(s_J)]\nonumber\\
    &=\frac12\qty(\pdv{S_J}{\phi_J})^2+\frac12\qty(\pdv{S_J}{\eta_J^i})^2+V(\phi_J)+\frac12(\partial_i\phi_J)(\partial^i\phi_J)+\frac12(\partial_i\eta^j)_J(\partial^i\eta_j)_J-\frac{1}{\kappa}\partial_i\eta^i_J,\label{qft_involutive_s1}
\end{align}
and
\begin{align}
    0&=\Tr\qty[\ket{X_{J+1}}\braket{X_{J+1}}{X_J}\bra{X_J}\hat{\mathcal H}_i(s_J)]\nonumber\\
    &=-\pdv{S_J}{\phi_J}\partial_i\phi_J-\pdv{S_J}{\eta_J^j}\qty(\partial_i\eta^j_J-\frac{\delta_i^j}{\kappa}).\label{qft_involutive_s2}
\end{align}
Then, we choose a set of operators that are used to integrate $S_J$. Let us take
\begin{equation}
    \mathcal H_{\rm evol}(s,\bfx)=\mathcal H(s,\bfx)+B^i(s,\bfx)\mathcal H_i(s,\bfx),
\end{equation}
where $B^i(s,\bfx)$ is an arbitrary function.
The characteristic curve for $\mathcal H_{\rm evol}$ is given by
\begin{align}
    \frac{\phi_{J+1}(\bfx)-\phi_J(\bfx)}{\delta s}&=-\alpha_J(\bfx)\pdv{S_J}{\phi_J(\bfx)}+\beta_J^i(\bfx)\partial_i\phi_J(\bfx),\label{qft_involutive_dphids}\\
    \frac{\eta^i_{J+1}(\bfx)-\eta^i_J(\bfx)}{\delta s}&=-\alpha_J(\bfx)\pdv{S_J}{\eta_J^i(\bfx)}+\beta_J^i(\bfx)\qty(\partial_j\eta_J^i(\bfx)-\frac{\delta_j^i}{\kappa}),\label{qft_involutive_detads}
\end{align}
where $\beta^i(s,\bfx)=\alpha(s,\bfx)B^i(s,\bfx)$. Then, $\mathcal H_{\rm evol}\approx0$ gives
\begin{align}
    \pdv{S_J}{\phi_J}\frac{\phi_{J+1}-\phi_J}{\delta s}+\pdv{S_J}{\eta_J^i}\frac{\eta^i_{J+1}-\eta^i_J}{\delta s}=2\alpha_J\mathcal V_J+\beta^i\qty[\pdv{S_J}{\phi_J}\partial_i\phi_J+\pdv{S_J}{\eta_J^j}\qty(\partial_i\eta^j_J-\frac{\delta_i^j}{\kappa})],
\end{align}
where the last term vanishes due to Eq.~\eqref{qft_involutive_s2}.

On the other hand, substituting Eqs.~\eqref{qft_involutive_dphids} and \eqref{qft_involutive_detads} into Eqs.~\eqref{qft_involutive_s1} and \eqref{qft_involutive_s2}, we obtain
\begin{align}
    &\alpha^2_J=\frac{\mathcal K_J}{-\mathcal V_J+i\epsilon},\\
    &\beta^i_J\mathcal M_{ij}^J=\frac{\phi_{J+1}-\phi_J}{\delta s}(\partial_j\phi_J)+\frac{\eta^i_{J+1}-\eta^i_J}{\delta s}\qty[(\partial_j\eta_i)_J-\frac{\delta_{ij}}{\kappa}],\label{eq_qft_involtive_mij}
\end{align}
where
\begin{align}
    \mathcal M_{ij}^J&=(\partial_i\phi_J)(\partial_j\phi_J)+(\partial_i\eta^k)_J(\partial_j\eta_k)_J-\frac{1}{\kappa}((\partial_i\eta_j)_J+(\partial_j\eta_i)_J)+\frac{\delta_{ij}}{\kappa^2},\\
    \mathcal K_J&=\frac12\qty(\frac{\phi_{J+1}-\phi_J}{\delta s}-\beta_J^i\partial_i\phi_J)^2+\frac12\qty(\frac{\eta^i_{J+1}-\eta^i_J}{\delta s}-\beta_J^j\partial_j\eta_J^i+\frac{\beta_J^i}{\kappa})^2,\\
    \mathcal V_J&=V(\phi_J)+\frac12(\partial_i\phi_J)(\partial^i\phi_J)+\frac12(\partial_i\eta^j)_J(\partial^i\eta_j)_J-\frac{1}{\kappa}\partial_i\eta^i_J.
\end{align}
These equations determine all of $\alpha_J$ and $\beta^i_J$'s if $\det \mathcal M^J\neq0$.

We integrate $S_J$ and obtain
\begin{align}
    \braket{X_{J+1}}{X_J}&\simeq1-\frac{i}{\hbar}\int\dd[3]{x}\qty[\pdv{S_J}{\phi_J(\bfx)}\frac{\phi_{J+1}(\bfx)-\phi_J(\bfx)}{\delta s}+\pdv{S_J}{\eta^i_J(\bfx)}\frac{\eta^i_{J+1}(\bfx)-\eta^i_J(\bfx)}{\delta s}]\delta s\nonumber\\
    &=1+\frac{2i}{\hbar}\int\dd[3]{x}\sqrt{\mathcal K_J(\bfx)}\sqrt{-\mathcal V_J(\bfx)}\delta s.
\end{align}
Then, the Green's function is obtained as
\begin{align}
    G(\phi_n,\eta_n,\phi_1,\eta_1)&=\int\mathcal D\phi\mathcal D\eta\exp\qty[\frac{i}{\hbar}\int\dd[3]{x}\int_{0}^{1}\dd{s}2\sqrt{\mathcal K(s,\bfx)}\sqrt{-\mathcal V(s,\bfx)}],
\end{align}
where the path integral is executed over the paths satisfying $\det \mathcal M\neq0$ and
\begin{align}
    \mathcal M_{ij}&=(\partial_i\phi)(\partial_j\phi)+(\partial_i\eta^k)(\partial_j\eta_k)-\frac{1}{\kappa}(\partial_i\eta_j+\partial_j\eta_i)+\frac{\delta_{ij}}{\kappa^2},\\
    \mathcal K&=\frac12\qty(\dv{\phi}{s}-\beta^i\partial_i\phi)^2+\frac12\qty(\dv{\eta^i}{s}-\beta^j\partial_j\eta^i+\frac{\beta^i}{\kappa})^2,\\
    \mathcal V&=V(\phi)+\frac12(\partial_i\phi)(\partial^i\phi)+\frac12(\partial_i\eta^j)(\partial^i\eta_j)-\frac{1}{\kappa}\partial_i\eta^i.
\end{align}

With the four-dimensional metric,
\begin{equation}
    \tilde g_{\mu\nu}\dd{x^\mu}\dd{x^\nu}=-(\alpha^2-\beta^i\beta_i)\dd{s}^2+2\beta_i\dd{s}\dd{x^i}+\dd{x^i}\dd{x_i},
\end{equation}
the exponent can be rewritten as
\begin{align}
    2\sqrt{\mathcal K}\sqrt{-\mathcal V}=\sqrt{-\tilde g}\qty[-\frac12\tilde g^{\mu\nu}(\partial_\mu\phi)(\partial_\nu\phi)-\frac12\tilde g^{\mu\nu}(\partial_\mu\eta^i)(\partial_\nu\eta_i)-V(\phi)+\frac{1}{\kappa}\tilde g^{\mu i}\partial_\mu\eta_i+\frac{\beta^i\beta_i}{2\alpha^2\kappa^2}].\label{eq_qft_involtive_lagrangian}
\end{align}
Ignoring the second term and the last term, it reproduces Eq.~\eqref{eq_qft_lagrangian} after a transformation of \eqref{eq_qft_metric_transform} with appropriate $\lambda^i$'s. The differences from the previous example are that there are no constraints on $\eta^i$'s and that the $(si)$-components of the metric are non-zero. We also observe that Eqs.~\eqref{eq_qft_eom_beta} and \eqref{eq_qft_involtive_mij} are similar. They differ by the $1/\kappa^2$ term, which is because of the last term of Eq.~\eqref{eq_qft_involtive_lagrangian}. 

\section{Quantum Gravity}\label{sec_gravity}
\setcounter{equation}{0}
In the previous examples, we took $\kappa\to0$ to get rid of the effects of auxiliary fields. However, in quantum gravity, we will see that they are not auxiliary fields, but gravitational degrees of freedom; the structure of the $\order{\kappa^2}$ terms is uniquely determined, and $\kappa$ acquires the meaning of the gravitational constant. In this section, we obtain the path integral formulas for a scalar field theory with gravity and for pure gravity. We also examine the equations of motion and the decoupling limit of gravity.
\subsection{Path Integral for Scalar Field Theory with Gravity}\label{subsec_scalar_gravity}
In the ADM formulation \cite{Arnowitt:1959ah}, the Hamiltonian of a scalar field with gravity is given by
\begin{align}
    H&=\lambda\pi_N+\lambda_i\pi^i_N+N\mathcal H+N_i\mathcal H^i+\partial_\mu\mathcal H_{\rm bdy}^\mu,\label{eq_total_hamiltonian}
\end{align}
where $N$, $N_i$, $\lambda$ and $\lambda_j$ are Lagrange multipliers and
\begin{align}
    \mathcal H&=\frac{1}{\sqrt{h}}\qty[2\kappa G_{ijkl}\pi^{ij}\pi^{kl}+\frac{1}{2}\pi_\phi^2]+\sqrt{h}\qty[-\frac{1}{2\kappa}{}^{(3)}\mathcal R+\frac{1}{2}h^{ij}(\partial_i\phi)(\partial_j\phi)+V(\phi)],\\
    \mathcal H^i&=(\partial^i\phi)\pi_\phi-2\sqrt{h}\nabla_j\frac{\pi^{ij}}{\sqrt{h}},\label{eq_momentum_const}\\
    \mathcal H_{\rm bdy}^t&=\pi^{ij}h_{ij},\label{eq_h_bdy1}\\
    \mathcal H_{\rm bdy}^i&=2\pi^{ij}N_j-\pi^{kl}h_{kl}N^i+\frac{\sqrt{h}}{\kappa}\partial^iN.\label{eq_h_bdy2}
\end{align}
Here, $\kappa=8\pi G$ with $G$ being Newton's constant, $\nabla_i$ denotes the three-dimensional covariant derivative, ${}^{(3)}\mathcal R$ is the three-dimensional Ricci scalar for $h_{ij}$, and
\begin{equation}
    G_{ijkl}=\frac{1}{2}\qty(h_{ik}h_{jl}+h_{il}h_{jk}-h_{ij}h_{kl}).
\end{equation}
The conjugates of $N$, $N_i$, $\phi$ and $h_{ij}$ are denoted by $\pi_N$, $\pi_N^i$, $\pi_\phi$ and $\pi^{ij}$, respectively. In Eq.~\eqref{eq_momentum_const}, we divided $\pi^{ij}$ by $\sqrt{h}$ to keep track of the density weight. 
We raise and lower the spacial indices with $h_{ij}$ and the summation is implicit.

Eq.~\eqref{eq_total_hamiltonian} is essentially a linear combination of constraints and the system is defined through
\begin{equation}
    \pi_N\approx0,~\pi_N^i\approx0,~\mathcal H\approx0,~\mathcal H^i\approx0.
\end{equation}
The third equation is called the Hamiltonian constraint, or the Wheeler-deWitt equation \cite{DeWitt:1967yk,Wheeler:1968iap}, and the last equation is called the momentum constraints.
As we can see, these constraints are similar to those of the previous sections, where we found the constraints by ourselves. In quantum gravity, however, they are given as requirements of the theory. The relation between these constraints and those in the previous section will be shown in Subsection \ref{subsection_decoupling}.

The Poisson brackets among the constraints are $\{\mathcal H[f],\mathcal H[g]\}_P\approx0$, $\{\mathcal H[f],\mathcal H_i[g^i]\}_P\not\approx0$ and $\{\mathcal H_i[f^i],\mathcal H_j[g^j]\}_P\not\approx0$ in the presence of the scalar sector\footnote{There are attempts to recover diffeomorphisms with an extended phase space \cite{Isham:1984sb,Isham:1984rz}.}. Thus, only $\mathcal H$ is involutive operators in this example.

The constraints can be rewritten in a more convenient form as
\begin{align}
    \mathcal H&=\frac{1}{2}\Pi_M\frac{\gamma^{MN}}{\sqrt{h}}\Pi_N+\sqrt{h}\mathcal V,\label{eq_qg_ham_const}\\
    \mathcal H_i&=(\partial_i\Phi^M)\Pi_M-2\partial_k h_{ij}\pi^{jk},\label{eq_qg_mon_const}
\end{align}
where $\Phi^{\phi}=\phi$, $\Phi^{(ij)}=h_{ij}$, $\Pi_\phi=\pi_\phi$, $\Pi_{(ij)}=\pi^{ij}$, and
\begin{align}
    &\gamma^{\phi\phi}=1,~
    \gamma^{\phi (ij)}=0,~
    \gamma^{(ij)(kl)}=4\kappa G_{ijkl},\\
    &\mathcal V=-\frac{1}{2\kappa}{}^{(3)}\mathcal R+\frac{1}{2}h^{ij}(\partial_i\phi)(\partial_j\phi)+V(\phi).\label{eq_def_cal_v}
\end{align}

Following the Wheeler-deWitt formulation, these constraints are solved after the canonical quantization of the ADM variables,
\begin{align}
    [\hat h_{ij}(\bfx),\hat \pi^{kl}(\bfy)]&=\frac{i\hbar}{2}\qty(\delta_{i}^k\delta_{j}^l+\delta_{j}^k\delta_{i}^l)\delta^3(\bfx-\bfy),~
    [\hat \phi(\bfx),\hat \pi_\phi(\bfy)]=i\hbar\delta^3(\bfx-\bfy),\\
    [\hat N(\bfx),\hat \pi_N(\bfy)]&=i\hbar\delta^3(\bfx-\bfy),~
    [\hat N_i(\bfx),\hat \pi_N^j(\bfy)]=i\hbar\delta_i^j\delta^3(\bfx-\bfy).
\end{align}
We choose the operator ordering of Eq.~\eqref{eq_qg_ham_const}, and let the left (right) momentum operators operate always on the left (right) side. This prescription avoids unpleasant divergences.

The procedure to obtain the path integral formula is the same as the previous sections.
Since we have $\pi_N\approx0$ and $\pi_N^i\approx0$, it is sufficient to consider only the evolution of $\hat\phi$ and $\hat h_{ij}$. Thus, we define a basis at $s=s_J$ as
\begin{equation}
    \ket{X_J}=\ket{\Phi_J}.
\end{equation}

From the Hamiltonian constraint, the characteristic curve is determined as
\begin{align}
    \frac{\Phi^M_{J+1}(\bfx)-\Phi^M_J(\bfx)}{\delta s}&=-\alpha_J(\bfx)\frac{\gamma_J^{MN}}{\sqrt{h_J}}\pdv{S_J}{\Phi^N_J(\bfx)}.
\end{align}
Here, we introduced only one time variable, $\delta t_{\bfx}=\alpha_J(\bfx)\delta s$, at each $\bfx$ just like in the previous sections. In the gravitational theory, this also means that all the particles at $\bfx$ feel the same metric, $g_{ss}(\bfx)=-\alpha^2$, but not their own.
Then, the Lagrange multiplier is solved as $\alpha^2=\mathcal K/(-\mathcal V)$, where
\begin{align}
    \mathcal K&=\frac12\gamma_{MN}\pdv{\Phi^M}{s}\pdv{\Phi^N}{s}\nonumber\\
    &=\frac{(\partial_s\phi)^2}{2}+\frac{G^{ijkl}}{8\kappa}(\partial_sh_{ij})(\partial_sh_{kl}),\label{eq_def_cal_k}
\end{align}
with
\begin{equation}
    G^{ijkl}=\frac{1}{2}\qty(h^{ik}h^{jl}+h^{il}h^{jk}-2h^{ij}h^{kl}).
\end{equation}
Here, a difference from the previous examples is that $\gamma^{MN}$ is not positive definite; the modes that are proportional to $h_{ij}$ give negative contributions, which is not avoided even after the momentum constraints are applied. Thus, we have four possibilities\footnote{The cases with $\mathcal K<0$ look bizarre, but such a case has been discussed in the context of the creation of a universe from nothing \cite{Vilenkin:1982de}. We do not go into the validity of this possibility.}: (i) $\mathcal K>0,\mathcal V<0$, (ii) $\mathcal K>0,\mathcal V>0$, (iii) $\mathcal K<0,\mathcal V<0$, and (iv) $\mathcal K<0,\mathcal V>0$. For (i), we choose $\alpha>0$, which is simply a definition of the positive frequency modes. For (ii) and (iii), $\alpha$ is pure imaginary and we need to select a branch so that the probability dumps in the tunneling region. However, for (iv), there is no proper choice since the positive frequency modes in (iv) are not directly connected to those in (i). In this paper, we choose $\alpha>0$ for (iv) for the simplicity of the expression. Then, $\alpha$ can be written as
\begin{equation}
    \alpha=\frac{\sqrt{\mathcal K}}{\sqrt{-\mathcal V}}.
\end{equation}

Following the previous procedure, we obtain the Green's function as
\begin{equation}
    G(\Phi_n,\Phi_1)=\int\mathcal D\phi\mathcal Dh_{ij}\delta_h\exp\qty[\frac{i}{\hbar}\int_0^1\dd{s}\int\dd[3]{x}2\sqrt{h}\sqrt{\mathcal K(s,\bfx)}\sqrt{-\mathcal V(s,\bfx)}],
\end{equation}
whose exponent agrees with \cite{Oshita:2021aux}.
Here, the path integral is executed over the paths that satisfy
\begin{equation}
    0=\frac{\sqrt{h}}{\alpha}\gamma_{MN}(\partial_i\Phi^M)(\partial_s\Phi^N)-\frac{1}{2\kappa}\partial_k \qty(h_{il}\sqrt{h}\frac{G^{klmn}}{\alpha}\partial_sh_{mn}),\label{eq_qg_const_path}
\end{equation}
which eliminates three degrees of freedom from $h_{ij}$.

As shown in \cite{Oshita:2021aux}, the exponent can be rewritten as
\begin{align}
    \mathcal L&=2\sqrt{h}\sqrt{\mathcal K}\sqrt{-\mathcal V}\nonumber\\
    &=\alpha\sqrt{h}\qty[\frac{G^{ijkl}}{8\kappa\alpha^2}(\partial_sh_{ij})(\partial_sh_{kl})+\frac{1}{2\kappa}{}^{(3)}\mathcal R-\frac{g^{\mu\nu}}{2}(\partial_\mu\phi)(\partial_\nu\phi)-V(\phi)]\nonumber\\
    &=\sqrt{-g}\qty[\frac{1}{2\kappa}\mathcal R-\frac{g^{\mu\nu}}{2}(\partial_\mu\phi)(\partial_\nu\phi)-V(\phi)+\frac{1}{\kappa}D_\mu(u^\rho D_\rho u^\mu-u^\mu D_\rho u^\rho)],\label{eq_qg_action}
\end{align}
where $D_\mu$ is the four-dimensional covariant derivative for $g_{\mu\nu}$, $\mathcal R$ is the four-dimensional Ricci scalar, and $u^\mu$ is the same as the one we defined in Eq.~\eqref{eq_def_u}. Here, the four-dimensional metric is defined by
\begin{equation}
    g_{\mu\nu}\dd{x^\mu} \dd{x^\nu}=-\alpha^2\dd{s}^2+h_{ij}\dd{x^i}\dd{x^j}.
\end{equation}

\subsection{Equations of Motion}
As in the example of quantum field theory, the momentum constraints are not Lorentz covariant and thus we use a symmetry to recover the covariance.
Let us consider the same transformation given in Eq.~\eqref{eq_qft_metric_transform}. Then, Eq.~\eqref{eq_qg_action} becomes
\begin{align}
    \mathcal L_{\boldsymbol\lambda}&=\mathcal L-\frac{G^{ijkl}}{2\kappa}\frac{\sqrt{h}}{\alpha}(\partial_sh_{ij})(\nabla_k\lambda_l)-\lambda^i\frac{\sqrt{h}}{\alpha}(\partial_s\phi)(\partial_i\phi)\nonumber\\
    &\hspace{3ex}+\frac{G^{ijkl}}{2\kappa}\frac{\sqrt{h}}{\alpha}(\nabla_i\lambda_j)(\nabla_k\lambda_l)+\frac{\sqrt{h}}{2\alpha}\lambda^i\lambda^j(\partial_i\phi)(\partial_j\phi),
\end{align}
which is identical to $\mathcal L$ after the coordinate transformation, $\partial_\mu\to(\Lambda^{-1})^{~\nu}_\mu\partial_\nu$.

Next, we rewrite the momentum constraints in a convenient form,
\begin{align}
    0&=\lambda_i\qty(\frac{\sqrt{h}}{\alpha}(\partial_s\phi)(\partial^i\phi)-\sqrt{h}\nabla_j\frac{G^{ijkl}}{2\kappa\alpha}\partial_sh_{kl})&\nonumber\\
    &=\frac{\sqrt{h}}{\alpha}\qty(\lambda^i(\partial_i\phi)(\partial_s\phi)+\frac{G^{ijkl}}{2\kappa}(\partial_sh_{ij})(\nabla_k\lambda_l))-\frac{1}{2\kappa}\partial_j\frac{\sqrt{h}}{\alpha}\lambda_iG^{ijkl}\partial_sh_{kl}.
\end{align}

Then, we introduce Lagrange multipliers, $\beta^i(s,\bfx)$, and minimize
\begin{align}
    S_{\boldsymbol\beta}&=\int\dd[4]{x}\qty[\mathcal L-2\beta_i\qty(\frac{\sqrt{h}}{\alpha}(\partial_s\phi)(\partial^i\phi)-\sqrt{h}\nabla_j\frac{G^{ijkl}}{2\kappa\alpha}\partial_sh_{kl})]\nonumber\\
    &=\int\dd[4]{x}\qty[\mathcal L_{-\boldsymbol\beta}-2\beta_i\qty(\frac{\sqrt{h}}{\alpha}(\partial_s\phi)(\partial^i\phi)-\sqrt{h}\nabla_j\frac{G^{ijkl}}{2\kappa\alpha}\partial_sh_{kl})]\nonumber\\
    &=\int\dd[4]{x}\mathcal L_{\boldsymbol\beta}+\frac{1}{\kappa}\int\dd[4]{x}\partial_j\frac{\sqrt{h}}{\alpha}\beta_iG^{ijkl}\partial_sh_{kl}.
\end{align}
The second term is a total derivative and does not affect the equations of motion. Variating $S_{\boldsymbol\beta}$ with respect to $\beta^i$, we obtain
\begin{equation}
    \delta S_{\boldsymbol\beta}=\sqrt{h}\qty[\qty(\frac{\beta^j}{\alpha}(\partial_j\phi)(\partial^i\phi)-\nabla_j\frac{G^{ijkl}}{\kappa\alpha}(\nabla_k\beta_l))-\qty(\frac{1}{\alpha}(\partial_s\phi)(\partial^i\phi)-\nabla_j\frac{G^{ijkl}}{2\kappa\alpha}(\partial_sh_{kl}))]\delta\beta^i,
\end{equation}
which determines $\beta^i$.

Variating $S_{\boldsymbol\beta}$ with respect to $\phi$, we obtain
\begin{equation}
    \delta S_{\boldsymbol\beta}=\sqrt{-\tilde g}\qty[\tilde D_\mu\partial^\mu\phi-V'(\phi)]\delta\phi.
\end{equation}
Similarly, variating $S_{\boldsymbol\beta}$ with respect to $h_{ij}$, we obtain
\begin{equation}
    \delta S_{\boldsymbol\beta}=\sqrt{-\tilde g}\qty[-\frac{1}{2\kappa}(G^{ij}-\beta^i\beta^jG^{ss})+\frac{1}{2}(T^{ij}-\beta^i\beta^jT^{ss})]\delta h_{ij},
\end{equation}
where
\begin{equation}
    G_{\mu\nu}=\tilde {\mathcal R}_{\mu\nu}-\frac{\tilde {\mathcal R}}{2}\tilde g_{\mu\nu},~T^{\mu\nu}=\frac{2}{\sqrt{-\tilde g}}\fdv{S_M}{\tilde g_{\mu\nu}},
\end{equation}
with
\begin{equation}
    S_M=\int\dd[4]{x}\sqrt{-\tilde g}\qty[-\frac{\tilde g^{\mu\nu}}{2}(\partial_\mu\phi)(\partial_\nu\phi)-V(\phi)].
\end{equation}
Here, $\tilde{\mathcal R}=\tilde g^{\mu\nu}\tilde{\mathcal R}_{\mu\nu}$, $\tilde{\mathcal R}_{\mu\nu}$ is the Ricci tensor for $\tilde g_{\mu\nu}$, and the metric is defined as
\begin{equation}
    \tilde g_{\mu\nu}\dd{x^\mu}\dd{x^\nu}=-(\alpha^2-\beta^i\beta_i)\dd{s}^2+2\beta_i\dd{s}\dd{x^i}+h_{ij}\dd{x^i}\dd{x^j}.\label{eq_metric_shift}
\end{equation}
As discussed in \cite{Gerlach:1969ph,Oshita:2021aux}, the Hamiltonian constraint gives the $(ss)$-component of Einstein's equation and the momentum constraints give the $(si)$-components of Einstein's equation.
Putting everything together, all the conventional equations of motion are reproduced.

\subsection{Decoupling Limit of Gravity}\label{subsection_decoupling}
Let us take the decoupling limit of gravity and see whether we obtain the constraints that we found in quantum field theory.

We expand the metric in Cartesian coordinates\footnote{For a system with spherical symmetry, see Appendix \ref{apx_o3}.},
\begin{align}
    h_{ij}&=\delta_{ij}+\sigma_{ij}(s,\bfx),
\end{align}
with  $\sigma_{ij}=\order{\kappa}$.
We take $\kappa\to0$ of Eqs.~\eqref{eq_def_cal_k} and \eqref{eq_def_cal_v} and obtain
\begin{align}
    \mathcal K&=\frac{1}{2}(\partial_s\phi)^2+\order{\kappa},\label{eq_decoup_cal_k}\\
    \mathcal V&=V(\phi)+\frac{1}{2}(\partial_i\phi)(\partial^i\phi)-\frac{1}{\kappa}\partial_i\qty[\frac12\partial_j\qty(\sigma^{ij}-\delta^{ij}\sigma_k^k)]+\order{\kappa}.\label{eq_decoup_cal_v}
\end{align}
As for Eq.~\eqref{eq_qg_const_path}, we obtain
\begin{equation}
    \partial_s\qty[\frac12\partial_j\qty(\sigma^{ij}-\delta^{ij}\sigma_k^k)]=\kappa(\partial^i\phi)(\partial_s\phi)+\frac{(\partial_j\alpha)}{\alpha}\partial_s\qty[\frac12\qty(\sigma^{ij}-\delta^{ij}\sigma_k^k)]+\order{\kappa^2},\label{eq_decoup_mom}
\end{equation}
which determines three of the six degrees of freedom of $\sigma_{ij}$.

Since we ignored higher order terms in $\kappa$, we cannot determine the remaining three degrees of freedom of $\sigma_{ij}$.
Let us evaluate their effects on the Green's function.
We decompose \cite{Lifshitz:1945du,Bardeen:1980kt}
\begin{equation}
    \frac12\sigma_{ij}=\sigma^S\delta_{ij}+\qty(\partial_i\partial_j-\frac{\delta_{ij}}{3}\partial_k\partial^k)\sigma^{\parallel}+(\partial_i\sigma^{\perp}_j+\partial_j\sigma^{\perp}_i)+\sigma_{ij}^T,
\end{equation}
where
\begin{equation}
    \partial^i\sigma^{\perp}_i=0,~\partial^i\sigma_{ij}^T=0,~\delta^{ij}\sigma_{ij}^T=0.
\end{equation}
Then, the left-hand side of Eq.~\eqref{eq_decoup_mom} is rewritten as
\begin{equation}
    \frac12\partial^j\qty(\sigma_{ij}-\delta_{ij}\sigma_k^k)=\partial_i\qty(\frac23\partial_k\partial^k\sigma^{\parallel}-2\sigma^S)+\partial_k\partial^k\sigma^{\perp}_i.
\end{equation}
On the other hand, the right-hand side of Eq.~\eqref{eq_decoup_mom} contains
\begin{equation}
    \frac12\qty(\sigma_{ij}-\delta_{ij}\sigma_k^k)=\qty(\frac23\partial_k\partial^k\sigma^{\parallel}-2\sigma^S)\delta_{ij}+\qty(\partial_i\sigma^{\perp}_j+\partial_j\sigma^{\perp}_i)+\sigma_{ij}^T+\qty(\partial_i\partial_j-\delta_{ij}\partial_k\partial^k)\sigma^{\parallel}.\label{eq_decoup_dsdidj}
\end{equation}
Here, the last two terms correspond to the three undetermined degrees of freedom. Let us see their effects on $\mathcal V$. From Eq.~\eqref{eq_decoup_mom}, we have
\begin{align}
    \partial_i\qty[\frac12\partial_j\qty(\sigma^{ij}-\delta^{ij}\sigma_k^k)]=\int\dd{s}\qty{\frac{\kappa}{\alpha}\partial_i\qty[\alpha(\partial^i\phi)(\partial_s\phi)]+\frac{1}{2\alpha}\qty(\partial_i\partial_j\alpha-\delta_{ij}\partial_k\partial^k\alpha)\sigma^{ij}+\order{\kappa^2}}.\label{eq_decoup_v_last}
\end{align}
From the equations of motion for the metric, we find
\begin{equation}
    \frac{1}{\alpha}\qty(\partial_i\partial_j\alpha-\delta_{ij}\partial_k\partial^k\alpha)=\order{\kappa}.
\end{equation}
Thus, the second term of Eq.~\eqref{eq_decoup_v_last} is $\order{\kappa^2}$ and we can ignore the effects of the undetermined gravitational degrees of freedom.

Let us compare the above results with those of the last section. Eqs.~\eqref{eq_decoup_cal_k} and \eqref{eq_decoup_cal_v} are the same as Eqs.~\eqref{eq_qft_k} and \eqref{eq_qft_v} if we identify $\eta^i=\partial_j(\sigma^{ij}-\delta^{ij}\sigma^k_k)/2$.
In Eq.~\eqref{eq_decoup_mom}, however, we have additional terms proportional to the derivatives of $\alpha$ compared with Eq.~\eqref{eq_qft_dsdeta}.
As we have mentioned in Section \ref{sec_qft}, this is due to the choice of the auxiliary fields; we could use $\partial_j\eta^{ij}$ instead of $\eta^i$ and define
\begin{align}
    \mathcal H&=\frac12\pi^2+\frac{\kappa}{2}\pi^{ij}\pi_{ij}+\frac12(\partial_i\phi)(\partial^i\phi)+V(\phi)-\frac{1}{\kappa}\partial_i\partial_j\eta^{ij},\\
    \mathcal H^i&=(\partial^i\phi)\pi-\partial_j\pi^{ij},
\end{align}
where $\pi^{ij}$ is the momentum conjugate of $\eta_{ij}$.
They represent the local energy conservation law and satisfy $\{\mathcal H[f],\mathcal H[g]\}_P\approx0$. If we identify $\eta^{ij}=(\sigma^{ij}-\delta^{ij}\sigma^k_k)/2$, they reproduce Eqs.~\eqref{eq_decoup_cal_k}, \eqref{eq_decoup_cal_v} and \eqref{eq_decoup_mom} in the limit of $\kappa\to0$.

From the above observation, the Hamiltonian constraint and the momentum constraints can be understood as the local energy conservation law in the decoupling limit of gravity. Thus, assuming that quantum field theory is obtained in the decoupling limit, the local energy conservation law seems to be a more reasonable starting point than the Schr\"odinger equation in quantum field theory.

\subsection{Path Integral for Pure Gravity}
Finally, we obtain the path integral formula for pure gravity. The Hamiltonian constraint and the momentum constraints are reduced to
\begin{align}
    \mathcal H&=\pi^{ij}\frac{2\kappa G_{ijkl}}{\sqrt{h}}\pi^{kl}-\frac{\sqrt{h}}{2\kappa}{}^{(3)}\mathcal R,\\
    \mathcal H^i&=-2\sqrt{h}\nabla_j\frac{\pi^{ij}}{\sqrt{h}}.
\end{align}
Unlike in the case with a scalar field, the constraint algebra closes \cite{diraclecture};
\begin{align}
    \{\mathcal H[f],\mathcal H[g]\}_P&=\mathcal H_i[f\partial^ig-g\partial^if]\approx0,\\
    \{\mathcal H[f],\mathcal H_i[g^i]\}_P&=\mathcal H[-g^i\partial_if]\approx0,\\
    \{\mathcal H_i[f^i],\mathcal H_j[g^j]\}_P&=\mathcal H_i[f^j\partial_jg^i-g^j\partial_jf^i]\approx0.
\end{align}

The procedure to obtain the path integral formula is the same as in Subsection \ref{subsec_more_ic}.
We first see the equations required for the integration of $S_J$. Here, $S_J$ is defined according to the previous sections.
The Hamiltonian constraint gives
\begin{equation}
    0=\frac{2\kappa G_{ijkl}^J}{\sqrt{h^J}}\pdv{S_J}{h_{ij}^J}\pdv{S_J}{h_{kl}^J}-\frac{\sqrt{h^J}}{2\kappa}{}^{(3)}\mathcal R^J,
\end{equation}
and the momentum constraints give
\begin{align}
    0&=2\sqrt{h^J}\nabla_j\frac{1}{\sqrt{h^J}}\pdv{S_J}{h_{ij}^J}.
\end{align}
We consider $0\approx\mathcal H_{\rm evol}=\mathcal H+B_i\mathcal H^i$, and obtain
\begin{align}
    0&=\frac{2\kappa G_{ijkl}^J}{\sqrt{h^J}}\pdv{S_J}{h_{ij}^J}\pdv{S_J}{h_{kl}^J}-\frac{\sqrt{h^J}}{2\kappa}{}^{(3)}\mathcal R^J+2B_i\sqrt{h^J}\nabla_j\frac{1}{\sqrt{h^J}}\pdv{S_J}{h_{ij}^J}\nonumber\\
    &=\frac{2\kappa G_{ijkl}^J}{\sqrt{h^J}}\pdv{S_J}{h_{ij}^J}\pdv{S_J}{h_{kl}^J}-\frac{\sqrt{h^J}}{2\kappa}{}^{(3)}\mathcal R^J-\frac{2}{\alpha_J}(\nabla_j\beta_i^J)\pdv{S_J}{h_{ij}^J}+\frac{2\sqrt{h^J}}{\alpha_J}\nabla_j\frac{\beta_i^J}{\sqrt{h^J}}\pdv{S_J}{h_{ij}^J},\label{eq_pure_grav_dsdh}
\end{align}
where we defined $B_i=\beta_i^J/\alpha_J$ and $\alpha_J$ is taken to be the same as the Lagrange multiplier of the characteristic curve. Since we will multiply $2\alpha_J$ before integrating this equation, the last term can be seen as total derivative. Thus, ignoring the last term, we obtain the characteristic curve as
\begin{align}
    \frac{h_{ij}^{J+1}-h_{ij}^J}{\delta s}&=-\frac{4\kappa\alpha_J}{\sqrt{h^J}}G^J_{ijkl}\pdv{S_J}{h_{ij}^J}+\nabla_i\beta_j^J+\nabla_j\beta_i^J.\label{eq_pure_grav_dhds}
\end{align}
Substituting it into Eq.~\eqref{eq_pure_grav_dsdh}, we obtain
\begin{equation}
    \pdv{S_J}{h_{ij}^J}\frac{h_{ij}^{J+1}-h_{ij}^J}{\delta s}=-\frac{\alpha_J\sqrt{h^J}}{\kappa}{}^{(3)}\mathcal R^J-\partial_i\mathcal J^i_J-2\beta_j^J\sqrt{h^J}\nabla_i\frac{1}{\sqrt{h^J}}\pdv{S_J}{h_{ij}^J},
\end{equation}
where
\begin{equation}
    \mathcal J^i=\frac{\beta_j}{2\kappa\alpha}G^{ijkl}(\partial_sh_{kl}-\nabla_k\beta_l-\nabla_l\beta_k).
\end{equation}
The last term vanishes because of the momentum constraints.

With Eq.~\eqref{eq_pure_grav_dhds}, the Hamiltonian constraint gives $\alpha=\sqrt{\mathcal K}/\sqrt{-\mathcal V}$ with
\begin{align}
    \mathcal K&=\frac{G^{ijkl}}{8\kappa}(\partial_sh_{ij}-\nabla_i\beta_j-\nabla_j\beta_i)(\partial_sh_{kl}-\nabla_k\beta_l-\nabla_l\beta_k),\\
    \mathcal V&=-\frac{1}{2\kappa}{}^{(3)}\mathcal R,
\end{align}
and the momentum constraints give
\begin{equation}
    0=\frac{\sqrt{h}}{\kappa}\nabla_j\frac{G^{ijkl}}{\alpha}\qty(\nabla_k\beta_l-\frac12\partial_sh_{kl}).
\end{equation}
Here, we do not discuss the existence of $\alpha$ and $\beta_i$ for arbitrary $h_{ij}$. If no solution exists, the corresponding path is simply not integrated in the path integral.

Using the above equations, we obtain the Green's function as
\begin{equation}
    G(\Phi_n,\Phi_1)=\int\mathcal D\phi\mathcal Dh_{ij}\exp\qty[\frac{i}{\hbar}\int_0^1\dd{s}\int\dd[3]{x}\qty(2\sqrt{h}\sqrt{\mathcal K}\sqrt{-\mathcal V}+\partial_i\mathcal J^i)].
\end{equation}
Here, the first term in the exponent is known as the Baierlein-Sharp-Wheeler action \cite{Baierlein:1962zz}, the path integral of which has been discussed in \cite{Carlini:1995gn}.
As for the second term, we can rewrite the exponent in the four-dimensional form as \cite{Oshita:2021aux}
\begin{equation}
    2\sqrt{h}\sqrt{\mathcal K}\sqrt{-\mathcal V}+\partial_i\mathcal J^i=\frac{\sqrt{-\tilde g}}{2\kappa}\tilde{\mathcal R}+\partial_\mu\mathcal H^\mu_{\rm bdy},
\end{equation}
where $\tilde g_{\mu\nu}$ is defined in Eq.~\eqref{eq_metric_shift}, $\tilde {\mathcal R}$ is its four-dimensional Ricci scalar, and $\mathcal H^\mu_{\rm bdy}$ is given by Eqs.~\eqref{eq_h_bdy1} and \eqref{eq_h_bdy2} with replacements, $N\to\alpha$ and $N_i\to\beta_i$. This result is in agreement with \cite{Tanaka:1992zw,Gen:1999gi}.

\section{Relation to the Conventional Path Integral}\label{sec_relation}
In the previous sections, we have obtained the path integral formulas, whose exponents contain square roots.
Since the exponents can become complex, our formulas naturally describe the mixed tunneling.
In special cases where $\alpha$ is constant, we have seen that our path integral is identical to the conventional one, but we have not discussed their relation for general $\alpha$.

Since we know that there are solutions to the Schr\"odinger equation that can be understood as the mixed tunneling, the conventional path integral should also be possible to describe the mixed tunneling. In this section, we obtain our path integral formula, more precisely the square root action, from the conventional one using a simple example.
\subsection{Conventional Path Integral}
Let us first briefly review the derivation of the conventional path integral.
We consider a system with Hamiltonian,
\begin{equation}
    H=\sum_{a=1}^N\frac{m}{2}\qty(\dv{x^a}{t})^2+V(X),
\end{equation}
where $X=\{x^1,x^2,\dots,x^N\}$.

We discretize time as $0=t_0>t_{-1}>t_{-2}>\cdots$ and define the basis at $t=t_{-J}$ as $\ket{X_{-J}}$, which is the eigenstate of $\hat x^a$.
Here, we use the same basis for all $J$, {\it i.e.} $\ket{X_{-J}}=\ket{X_{-J+1}}$, unlike in our formalism. We also define the momentum space, $\ket{P_{-J}}$, in the conventional way.
The Green's function from $t_i=-\infty$ to $t_f=0$ with $x(t_i)=x_i$ and $x(t_f)=x_f$ is then calculated as
\begin{align}
    &G_{\rm conv}(x_f,x_i;0,-\infty)\nonumber\\
    &=\prod_{J=1}^{\infty}\int\dd{X_{-J}}\bra{X_{-J}}e^{-\frac{i}{\hbar}H(\hat X,\hat P)\Delta t_{-J}}\ket{X_{-J+1}}\nonumber\\
    &\simeq\prod_{J=1}^{\infty}\int\dd{X_{-J}}\int\dd{P_{-J}}\bra{X_{-J}}\qty(1-\frac{i}{\hbar}H(X_{-J},P_{-J})\Delta t_{-J})\ket{P_{-J}}\bra{P_{-J}}\ket{X_{-J+1}}\nonumber\\
    &\simeq\prod_{J=1}^{\infty}\int\dd{X_{-J}}\int\dd{P_{-J}}e^{-\frac{i}{\hbar}H(X_{-J},P_{-J})\Delta t_{-J}}e^{-\frac{i}{\hbar}P_{-J}\cdot(X_{-J+1}-X_{-J})}\nonumber\\
    &=\prod_{J=1}^{\infty} \int\dd{X_{-J}}\qty(\frac{m}{2\pi i\hbar\Delta t_{-J}})^{N/2}e^{\frac{i}{\hbar}\qty[\sum_{a=1}^N\frac{m}{2}\qty(\frac{x^a_{-J+1}-x^a_{-J}}{\Delta t_{-J}})^2-V(X_{-J})]\Delta t_{-J}}\nonumber\\
    &=\int\mathcal DX_{\rm conv}e^{\frac{i}{\hbar}\int_{-\infty}^0\dd{t}L},\label{eq_conv_path_int}
\end{align}
where we took $\Delta t_{-J}\to0$ in the last line. Here, we defined $\Delta t_{-J}=t_{-J+1}-t_{-J}$,
\begin{equation}
    \int\dd{P_{-J}}=\prod_{a=1}^N\int\frac{\dd{p_a^{-J}}}{2\pi\hbar},
\end{equation}
and
\begin{equation}
    L=\sum_{a=1}^N\frac{m}{2}\qty(\dv{x^a}{t})^2-V(X).
\end{equation}
The last two lines of Eq.~\eqref{eq_conv_path_int} are the definition of the conventional path integral and its measure.
\subsection{Our Path Integral}
Let us first discuss a variation of the above formulation. We find
\begin{equation}
    0=\bra{X_{-J}}\qty[(\hat x_\parallel)^n,(\hat p^a_\perp)^m]\ket{X_{-J+1}}=\qty[(X_{-J})_\parallel^n-(X_{-J+1})_\parallel^n]\bra{X_{-J}}(\hat p^a_\perp)^m\ket{X_{-J+1}},
\end{equation}
where $\hat x_\parallel$ is the position operator parallel to $(X_{-J+1}-X_{-J})$ and $\hat p^a_\perp$ is one of the momentum operators perpendicular to it. For $X_{-J}\neq X_{-J+1}$, it gives $\bra{X_{-J}}(\hat p^a_\perp)^m\ket{X_{-J+1}}=0$.

The same result can be obtained by directly evaluating the momentum integral in Eq.~\eqref{eq_conv_path_int}.
In the third line, the momentum integral has to be evaluated with a cutoff because otherwise the exponential in the second line could not be expanded and also the integral would result in $\prod_a\delta(x^a_{-J+1}-x^a_{-J})$. Thus, we introduce the momentum cutoff at $p_{\rm max}\lesssim\sqrt{2m\hbar/\Delta t_{-J}}$. For general $n$-dimensional vectors, $p$ and $\Delta x$, we have
\begin{align}
    \int\frac{\dd[n]{p}}{(2\pi\hbar)^n}e^{-\frac{|p|^2}{2p_{\max}^2}}&=\qty(\frac{p_{\max}}{\sqrt{2\pi}\hbar})^{n},\\
    \int\frac{\dd[n]{p}}{(2\pi\hbar)^n}|p|^2e^{-\frac{i}{\hbar}p\cdot\Delta x-\frac{|p|^2}{2p_{\max}^2}}&=-(2\pi\hbar)^2\qty(\frac{p_{\max}}{\sqrt{2\pi}\hbar})^{n+4}\qty(|\Delta x|^2-\frac{n\hbar^2}{p_{\max}^2})e^{-\frac{p_{\max}^2|\Delta x|^2}{2\hbar^2}}.
\end{align}
Using these, we can see $\bra{X_{-J}}\sum_{a=1}^N(\hat p^a)^2\ket{X_{-J+1}}\simeq\bra{X_{-J}}\hat p_\parallel^2\ket{X_{-J+1}}$ for $p_{\max}\gg1/|\Delta x|$. Here, $\hat p_\parallel$ is the momentum operator parallel to $(X_{-J+1}-X_{-J})$.

The above observations motivate us to take $\bra{X_{-J}}(\hat p^a_\perp)^2\ket{X_{-J+1}}=0$ in the third line of Eq.~\eqref{eq_conv_path_int}.
Notice that this does not conflict with the equations of motion, ${\dd{\hat x^a}}/{\dd{t}}=i[\hat H,\hat x^a]/\hbar$, which implies $\hat p^a\propto {\dd{\hat x^a}}/{\dd{t}}$.
Then, the path integral formula becomes
\begin{align}
    &G_{\rm conv'}(x_f,x_i;0,-\infty)\nonumber\\
    &\simeq\prod_{J=1}^{\infty}\int\dd{X_{-J}}\int\dd{P_{-J}}e^{-\frac{i}{\hbar}H(X_{-J},P^\parallel_{-J})\Delta t_{-J}}e^{-\frac{i}{\hbar}P_{-J}\cdot(X_{-J+1}-X_{-J})-\frac{P_{-J}\cdot P_{-J}}{2p_{\max}^2}}\nonumber\\
    &\simeq\prod_{J=1}^{\infty} \int\dd{X_{-J}}\sqrt{\frac{m}{2\pi i\hbar\Delta t_{-J}}}\qty(\frac{p_{\max}}{\sqrt{2\pi}\hbar})^{N-1}e^{\frac{i}{\hbar}\qty[\sum_{a=1}^N\frac{m}{2}\qty(\frac{x^a_{-J+1}-x^a_{-J}}{\Delta t_{-J}})^2-V(X_{-J})]\Delta t_{-J}}\nonumber\\
    &=G_{\rm conv}(x_f,x_i;0,-\infty)\prod_{J=1}^{\infty}\qty(\sqrt{\frac{\Delta t_{-J}}{m\hbar}}p_{\rm max})^{N-1}.
\end{align}
Here, $P^\parallel_{-J}$ is $P_{-J}$ with only the parallel component, the dot product indicates the sum over $a$ and we swapped\footnote{This is implicitly used also in Eq.~\eqref{eq_conv_path_int} and is justified by explicitly showing that $\Delta t_{-J}^2$ corrections vanish as $\Delta t_{-J}\to0$. Since it is a standard discussion, we omit the proof here.} the order of $\Delta t_{-J}\to0$ and $p_{\max}\to\infty$.
Notice that $G_{\rm conv}$ and $G_{\rm conv'}$ only differ by the overall factor. Since we are often interested in the ratio of the Green's functions, the overall factor cancels out and the two formulations give the same result. However, for the discussion below, the measure of $G_{\rm conv'}$ is more reasonable.

Let us see that the square root action can be obtained from the conventional path integral.
Instead of parameterizing a path by $\{x^a_{-J}\}$, we can consider a different parameterization, $\{\sigma^a_{-J}\}$ and $\{\Delta t_{-J}\}$:
\begin{align}
    x^a(t_{-J})&=\sum_{K=1}^J\sigma^a_{-K}\Delta r,\\
    t_{-J}&=-\sum_{K=1}^J\Delta t_{-K},
\end{align}
where $\Delta r>0$ is a small parameter and $\Delta t_{-J}>0$ can become arbitrarily large. Here, $\{*\}$ denotes a set of the variables and
\begin{equation}
    \sum_{a=1}^N\qty(\sigma_{-J}^a)^2=1,
\end{equation}
where $J=1,\dots,N_\sigma$ and $N_\sigma$ can be different for different paths, and $x^a(t)$ is linearly interpolated.
Notice that this is an injection mapping from $\{\sigma_{-J}^a\}$ and $\{\Delta t_{-J}\}$ to the space of paths, just like the mapping from $\{x_{-J}^a\}$. In the limit of $\Delta r\to0$, the mapping spans all the possible paths.

Let us obtain the Jacobian for the change of the variables from $\{x^a_{-J}\}$ to $\{\sigma_{-J}^a\}$ and $\{\Delta t_{-J}\}$.
Let us define
\begin{equation}
    r_{-J}^2=\sum_a(x_{-J+1}^a-x_{-J}^a)^2,
\end{equation}
Then, we rewrite
\begin{align}
    \int\dd{X_{-J}}=\int\dd{\Omega_{-J}}\int_0^\infty\dd{r_{-J}}r_{-J}^{N-1},
\end{align}
where $\int\dd{\Omega_{-J}}$ is the integral of the angular part.
The angular integral is understood as the integral over $\{\sigma_{-J}^a\}$.
To find the Jacobian from $\{r_{-J}\}$ to $\{\Delta t_{-J}\}$, let us examine the conventional path integral measure sampled at $\{t_{-J}\}$.
Since we have
\begin{equation}
    \pdv{r_{-J}}{\Delta t_{-K}}\simeq
    \begin{cases}
        -\frac{\Delta r}{\Delta t_{-J}}+\frac{\Delta r}{\Delta t_{-J+1}}(\sigma_{-J}\cdot\sigma_{-J+1})&(K< J)\\
        -\frac{\Delta r}{\Delta t_{-J}}&(K= J)\\
        0&(K>J)
    \end{cases},
\end{equation}
the Jacobian is calculated as
\begin{equation}
    |\mathcal J|=\prod_{J=1}^{N_\sigma}\frac{\Delta r}{\Delta t_{-J}}.
\end{equation}
Then, we obtain
\begin{align}
    &G_{\rm conv'}(x_f,x_i;0,-\infty)\nonumber\\
    &\simeq\sum_{N_\sigma=1}^\infty\prod_{J=1}^{N_\sigma}\int\dd{\Omega_{-J}}\qty(\frac{p_{\max}}{\sqrt{2\pi}\hbar})^{N-1}\int_0^\infty\dd{\Delta t_{-J}}\frac{\Delta r^{N}}{\Delta t_{-J}}\sqrt{\frac{m}{2\pi i\hbar\Delta t_{-J}}}e^{\frac{i}{\hbar}\qty[\frac{m}{2}\qty(\frac{\Delta r}{\Delta t_{-J}})^2-V(X_{-J})]\Delta t_{-J}}\nonumber\\
    &=\sum_{N_\sigma=1}^\infty\prod_{J=1}^{N_\sigma}\int\dd{\Omega_{-J}}\qty(\frac{p_{\max}\Delta r}{\sqrt{2\pi}\hbar})^{N-1}\exp[\frac{i}{\hbar}2\sqrt{\frac{m}{2}\qty(\frac{\Delta r}{\Delta s})^2}\sqrt{-V(X_{-J})}\Delta s]\nonumber\\
    &=\int\mathcal D\Omega \exp[\frac{i}{\hbar}\int_0^1\dd{s}2\sqrt{\frac{m}{2}\qty(\dv{X}{s}\cdot\dv{X}{s})}\sqrt{-V(X)}].
\end{align}
In the last line, we took $p_{\max}\to\infty$ and $\Delta r\to0$ with keeping $p_{\max}\Delta r\simeq\sqrt{2\pi}\hbar$, which is reasonable because the resolution of the position space with the momentum cutoff is roughly given by $\hbar/p_{\max}$. The path integral measure $\int\mathcal D\Omega$ is the same as the one in the previous sections up to an overall factor.
It is a virtue of our path integral formula that the path integral measure is not singular in the continuous limit. Since the expression is invariant under the change of $\Delta s$, we rescaled $s$ so that $s\in[0,1]$.

Now, we move on to the mixed tunneling. We consider a partially separable system, where
we have non-overlapping subsets $X^A\subset X$ and the Hamiltonian is separated as
\begin{equation}
    H(\hat X,\hat P)=\sum_AH_A(\hat X^A,\hat P^A).
\end{equation}
Since $[H_A(\hat X^A,\hat P^A),H_B(\hat X^B,\hat P^B)]=0$, we can evolve the system separately as
\begin{align}
    &G_{\rm conv''}(x_f,x_i;0,-\infty)\nonumber\\
    &=\prod_{J=1}^{\infty}\prod_A\int\dd{X_{-J}^A}\int\dd{P_{-J}^A}e^{-\frac{i}{\hbar}H^A(X^A_{-J},P^A_{-J})\Delta t_{-J}^A}e^{-\frac{i}{\hbar}P^A_{-J}\cdot(X^A_{-J+1}-X^A_{-J})}\nonumber\\
    &=\int\mathcal D\Omega \exp[\frac{i}{\hbar}\sum_A\int_0^1\dd{s}2\sqrt{\frac{m}{2}\qty(\dv{X^A}{s}\cdot\dv{X^A}{s})}\sqrt{-V_A(X^A)}],\label{eq_conv_pp}
\end{align}
where $V_A(X^A)$ is the potential term in $H_A$ and we took different time durations for different $H_A$. When each of $X^A$'s has only one element, the result agrees with our path integral formula in Section \ref{sec_separable}. We can see that the exponent can become complex even though we started from the conventional path integral formula, whose exponent is pure imaginary.

Once we consider interactions between $X^A$ and $X^B$, the Hamiltonian is not separable and we cannot evolve $X^A$ and $X^B$ separately in a naive way.
On the other hand, we know that this problem has already been overcome in the previous sections in a different manner.
Let us reinterpret what we have done in Section \ref{sec_non_separable}. For a small $\Delta r$, we can ignore the energy transfer between $A$ and $B$ since it is proportional to the change of $X^A-X^B$. It enables us to define the local energy associated with $X^A$. Then, we can uniquely split the Hamiltonian as $H=\sum_AH_A$ with the help of additional variables, $\eta^A$, which store the information about the local energy. We have seen that the proper evolution of $\eta^A$ makes $\{H_A,H_B\}_P\approx0$ on the constrained path. Thus, we can treat the system just as a separable system and evolve $X^A$ independently of $X^B$. Then, the time spent changing $X^A$ can be different from the time spent changing $X^B$ and this results in the mixed tunneling. Since it would be hard to derive and justify all of these in this fashion, including the alternative path integral measure we have discussed, we stop here and leave the explicit derivation elsewhere.

\section{Summary and Discussions}\label{sec_summary}
We presented a new path integral formalism that can be applied for mixed tunneling, polychronic tunneling and quantum gravity. Our formulation is based on the local energy conservation law instead of the explicit time evolution through the Schr\"odinger equation. We obtain the path integral formulas that have square root actions, which are generally complex. Although they look different from the conventional one, we have seen that our formula is identical to the conventional one when $\alpha$ is constant. In addition, we have seen that our formula can be obtained from the conventional one in the simplest case.
We also found that the conventional equations of motion are reproduced even though the constraints are not Lorentz covariant.

It is worth noting that the time-independent Schr\"odinger equation does not give enough information to determine the exponent of the wave function for a many-body system.
The main point of this paper is that we can provide the missing information through the local energy conservation law.
In quantum gravity, this is the very definition of the theory. In fact, we found that the path integral formula for quantum gravity reproduces that of quantum field theory in the decoupling limit of gravity, provided that the auxiliary fields are appropriately chosen.

In our formulation, global time disappears from the theory.
This has been known as the problem of time in quantum gravity, which becomes a common problem in our formulation. Our solution to this problem is similar to that previously discussed as the emergent time in the WKB approximation.
As we have seen in the previous sections, the concept of time first appears when we relate the deformation parameter, $s$, to the local time variable through $\alpha^a$. The local time, $\dd{t^a}$, is defined by how much a particle has moved, $|\dd{x^a}|$, relative to its velocity, $\sqrt{2m(\mathcal E-\delta_{a0}V(x^a))}/m$. 
With this definition, the local time becomes complex and evolves towards positive real or negative imaginary depending on whether the evolution is Lorentzian or Euclidean.

In \cite{Oshita:2021aux}, a new tunneling process, polychronic tunneling, has been proposed in quantum gravity. They have numerically shown that it typically enhances the decay rate of a meta-stable vacuum by many orders of magnitude. They also discussed the existence of polychronic tunneling in the decoupling regime of gravity, but the corresponding formulation without gravity has not been available. This paper provides the formulation for quantum field theory without gravity and supports their results.

As mixed tunneling has been studied for decades in various fields, there already exist various experimental results, which make our formalism testable. In addition, several experiments to measure the tunneling time have come out using strong field tunneling ionization \cite{PhysRevLett.119.023201} and ultracold atoms \cite{Ramos:2019one,Spierings:2021xiq}. These can also be used to test our formalism since they are essentially the system of a tunneling particle coupled to a clock.
For quantum field theory, the thermal tunneling rate of a spin chain has recently been measured using superfluid atomic systems \cite{zenesini2023observation}. The effective local temperature of the system is approaching the quantum tunneling regime and the mixed tunneling of spin chains may be observed in the future.
Furthermore, comparisons can be made with the other techniques based on the complex trajectories, the adiabatic approximation and the Huygens-Fresnel principle.
The studies of these directions will be provided elsewhere.

It would also be interesting to discuss the second quantization in our formalism, although the locality of the constraints is not very comfortable when we Fourier-transform fields. For example, in the conventional path integral, we evaluate loop integrals with analytic continuation because otherwise the integrals are not well-defined. However, our formula will automatically continue the integrals analytically when the required energy exceeds the locally available energy. The consistency with the conventional loop integrals is therefore non-trivial and needs to be checked.
It may also be interesting to discuss the optical theorem in our framework since our formula has a predefined analytic structure.
To see the consistency of our formalism further, we need to include fermions and gauge fields. Once it is done, an interesting direction would be to study the strong CP phase since time is emergent and the total derivative terms are not arbitrary in our formulation. Another direction would be to compare the results of our formula with the conventional one using quantum simulations of lattice gauge theory.

\begin{acknowledgments}
    Y.S.~is supported by the US-Israeli Binational Science Foundation (grant No. 2020220) and the Israel Science Foundation (grant No. 1818/22).
\end{acknowledgments}

\appendix

\section{Method of Characteristics and Complex Exponent}\label{apx_char}
We give a review on the method of characteristics in terms of Lagrange multipliers and discuss its relation to the mixed tunneling.

Let us consider the non-linear partial differential equation given by
\begin{equation}
    0=H(x^1,\dots,x^N,p_1,\dots,p_N),\label{eq_apx_dif_eq}
\end{equation}
with
\begin{equation}
    p_a=\pdv{S}{x^a}.\label{eq_apx_p_def}
\end{equation}
In particular, we are interested in
\begin{equation}
    H=\sum_{ab}\frac12\gamma^{ab}p_ap_b+V,\label{eq_apx_h_of_interest}
\end{equation}
where $\gamma^{ab}$ and $V$ are functions of $\{x^1,\dots,x^N\}$. We assume $\gamma^{ab}$ is an invertible symmetric matrix, $\sum_b\gamma^{ab}\gamma_{bc}=\delta_{ac}$.

Suppose we have a solution, $S(x^1,\dots,x^N)$, to Eq.~\eqref{eq_apx_dif_eq}. We consider an arbitrary curve, $(x^1(s),\dots,x^N(s))$, on the domain of $S$.
Along the curve, we have
\begin{align}
    0=\dd{H}=\sum_a\pdv{H}{x^a}\dd{x^a}+\sum_a\pdv{H}{p_a}\dd{p_a}.\label{eq_apx_dh}
\end{align}
Here, $\dd{x^a}$'s are understood as independent variables since we can consider other curves passing through the same point. However, $\dd{p_a}$'s are not independent due to the constraints of \eqref{eq_apx_p_def}. Since $p_a$ also becomes a function of $s$ along the curve, we have
\begin{equation}
    \dv{p_a}{s}\dd{x^a}=\dv{p_a}{s}\dv{x^a}{s}\dd{s}=\dv{x^a}{s}\dd{p_a},\label{eq_apx_dpdsdx}
\end{equation}
for each $a$.

For the moment, we consider only the following combination of the constraints,
\begin{equation}
    0=\sum_a\qty(\dv{p_a}{s}\dd{x^a}-\dv{x^a}{s}\dd{p_a}).
\end{equation}
Under this constraint, Eq.~\eqref{eq_apx_dh} becomes
\begin{equation}
    0=\sum_a\qty(\pdv{H}{x^a}+\frac{1}{\alpha}\dv{p_a}{s})\dd{x^a}+\sum_a\qty(\pdv{H}{p_a}-\frac{1}{\alpha}\dv{x^a}{s})\dd{p^a},\label{eq_apx_alpha_dh}
\end{equation}
where $\alpha^{-1}(s)$ is the Lagrange multiplier.
Although $\dd{x^a}$ and $\dd{p_a}$ are still not fully independent, we shall treat them as independent variables. This results in additional constraints on the solution, which we will discuss later. We obtain
\begin{align}
    \pdv{H}{p_a}&=\frac{1}{\alpha}\dv{x^a}{s},\label{eq_apx_LC1}\\
    \pdv{H}{x^a}&=-\frac{1}{\alpha}\dv{p_a}{s}.\label{eq_apx_LC2}
\end{align}
Using Eq.~\eqref{eq_apx_LC1}, we also have
\begin{equation}
    \dv{S}{s}=\sum_a\pdv{S}{x^a}\dv{x^a}{s}=\alpha\sum_a\pdv{H}{p_a}p_a.\label{eq_apx_LC3}
\end{equation}
In the symmetric form, the above equations are called the Lagrange-Charpit equations,
\begin{equation}
    \frac{\dd{x^a}}{\pdv{H}{p_a}}=-\frac{\dd{p_b}}{\pdv{H}{x^b}}=\frac{\dd{S}}{\sum_c\pdv{H}{p_c}p_c}.
\end{equation}

The method of characteristics is the method to construct $S$ using the Lagrange-Charpit equations. First, Eqs.~\eqref{eq_apx_LC1} and \eqref{eq_apx_LC2} defines a curve, which is called as the characteristic curve. It can be expanded around $s=0$ as
\begin{align}
    x^a(\delta s)&=x^a(0)+\dv{x^a}{s}(0)\delta s+\frac12\dv[2]{x^a}{s}(0)\delta s^2+\cdots\nonumber\\
    &=x^a(0)+\alpha\pdv{H}{p_a}\delta s+\frac{\alpha^2}{2}\qty[\frac{\dv{\alpha}{s}}{\alpha^2}\pdv{H}{p_a}+\sum_b\qty(\pdv{H}{p_b}\pdv{H}{p_a}{x^b}-\pdv{H}{x^b}\pdv{H}{p_a}{p_b})]\delta s^2+\cdots.
\end{align}
Along this curve, we can solve $S$ using Eq.~\eqref{eq_apx_LC3}. Around $s=0$, the solution is given by
\begin{align}
    S(\delta s)&=S(0)+\dv{S}{s}(0)\delta s+\frac12\dv[2]{S}{s}(0)\delta s^2+\cdots\nonumber\\
    &=S(0)+\alpha\sum_a\pdv{H}{p_a}p_a\delta s\nonumber\\
    &\hspace{3ex}+\frac{\alpha^2}{2}\sum_a\qty[\frac{\dv{\alpha}{s}}{\alpha^2}\pdv{H}{p_a}p_a+\sum_b\qty(\pdv{H}{p_b}\pdv{H}{p_a}{x^b}-\pdv{H}{x^b}\pdv{H}{p_a}{p_b})p_a-\pdv{H}{p_a}\pdv{H}{x^a}]\delta s^2+\cdots.
\end{align}

In particular, if $H$ has the form of Eq.~\eqref{eq_apx_h_of_interest}, we can solve Eq.~\eqref{eq_apx_LC1} for $p_a$ as
\begin{equation}
    p_a=\sum_b\frac{\gamma_{ab}}{\alpha}\dv{x^b}{s}.
\end{equation}
Substituting it to Eq.~\eqref{eq_apx_dif_eq}, we obtain
\begin{equation}
    \alpha^2=\frac{\frac12\sum_{ab}\gamma_{ab}\dv{x^a}{s}\dv{x^b}{s}}{-V}.
\end{equation}
In the linear order in $\delta s$, the characteristic curve is given by
\begin{equation}
    x^a(\delta s)\simeq x^a(0)+\alpha\sum_b\gamma^{ab}\pdv{S}{x^b}\delta s,
\end{equation}
and the solution is given by
\begin{align}
    S(\delta s)&\simeq S(0)+\frac{1}{\alpha}\sum_{ab}\gamma_{ab}\dv{x^a}{s}\dv{x^b}{s}\delta s\nonumber\\
    &=S(0)-2\alpha V\delta s.
\end{align}

There are some caveats to the method of characteristics.
First, if we take a different $\dd{x^a}/\dd{s}$, the solution is different in general because it corresponds to choosing a different $\partial S/\partial x^a$, which is determined by each solution. Second, the method constructs a solution in the vicinity of the characteristic curve, but the solution is not necessarily defined on all domain. It also does not guarantee that $S$ is single-valued. Finally, the method cannot construct all possible solutions as we have seen in Subsection \ref{subsec_lecl}. The last point is discussed in more detail below.

In the above derivation, we have treated $\dd{x^a}$ and $\dd{p_a}$ independently even though they are not fully independent.
To keep $S$ as general as possible, let us impose all of Eq.~\eqref{eq_apx_dpdsdx}. Introducing $N$ Lagrange multipliers, $\alpha^a$, we have
\begin{equation}
    0=\sum_a\qty(\pdv{H}{x^a}+\frac{1}{\alpha^a}\dv{p_a}{s})\dd{x^a}+\sum_a\qty(\pdv{H}{p^a}-\frac{1}{\alpha^a}\dv{x^a}{s})\dd{p^a}.
\end{equation}
Now, $\dd{x^a}$ and $\dd{p_a}$ are independent and we obtain
\begin{align}
    \pdv{H}{p_a}&=\frac{1}{\alpha^a}\dv{x^a}{s},\\
    \pdv{H}{x^a}&=-\frac{1}{\alpha^a}\dv{p_a}{s}.
\end{align}
However, we cannot determine all of $\alpha^a$'s only with Eq.~\eqref{eq_apx_dif_eq}. This happened because we extended the space of solution. The characteristic curve is not sufficient to specify a single solution and further information needs to be provided to integrate $S$.

In the standard version of the method of characteristics, we have additional constraints, $\alpha=\alpha^1=\dots=\alpha^N$, and this picks up a single solution. Let us interpret the physical meaning of these.
Before these constraints are imposed, each particle has its own time variable defined by
\begin{equation}
    \dd{t^a}=\alpha^a\dd{s}=\frac{\dd{x^a}}{v^a},
\end{equation}
with
\begin{equation}
    v^a=\pdv{H}{p^a}.
\end{equation}
Here, $\dd{x^a}$ is the distance actually traveled and $v^a$ is the distance that is expected to be traveled in unit time.
Then, the constraints imply the existence of global time, $t$,
\begin{equation}
    \dd{t}=\dd{t^1}=\dots=\dd{t^N}.\label{eq_apx_global_time}
\end{equation}
This means that we can determine the momentum and the kinetic energy of $x^a$ simply by $\dd{x^a}$. In general relativity, this corresponds to choosing a gauge, or equivalently a foliation, so that the time component of the metric becomes $g_{tt}=-1$. However, this is possible only when $\alpha^a$'s are all real numbers. For the mixed tunneling, some of them are pure imaginary and we do not have global time.

In our formulation of the mixed tunneling, we do not impose Eq.~\eqref{eq_apx_global_time} and keep the solution as general as possible. Instead, we determine $\alpha^a$ by the local energy conservation law.

\section{Interference and Quantum Tunneling}\label{apx_boundary_2d}
We demonstrate the importance of interference in the quantum tunneling in a many-body system.
Let us consider two particles whose positions are $x$ and $y$. The Schr\"odinger equation is given by
\begin{equation}
    \qty[-\frac{1}{2m}\nabla^2+V_{\rm step}(x)]\psi(x,y)=E\psi(x,y),
\end{equation}
where $\nabla=(\partial_x,\partial_y)$ and
\begin{equation}
    V_{\rm step}(x)=
    \begin{cases}
        V_0&(x\geq0)\\
        0&(x<0)
    \end{cases}.
\end{equation}
We consider the incoming wave of
\begin{align}
    \psi_{\rm in}(x<0,y)&=e^{ip_xx+ip_yy},
\end{align}
where
\begin{equation}
    \frac{p_x^2}{2m}<V_0<\frac{p_x^2+p_y^2}{2m}=E.
\end{equation}
Then, the total kinetic energy of the two particles is higher than the potential barrier, as in the example in Section \ref{sec_separable}.

Let us see that the Huygens principle reproduces the correct exponent for $x\geq0$ even though the individual spherical wavelet has a frequency high enough to penetrate the potential. In general, the wave function at the boundary can be written as
\begin{align}
    \psi_{x\geq0}(x=0,y)&=Ae^{ip_yy},\label{eq_bdy_a}\\
    \partial_x\psi_{x\geq0}(x=0,y)&=Be^{ip_yy},
\end{align}
where $A$ and $B$ are complex coefficients.

We prepare the Green's function for $x\geq0$,
\begin{equation}
    \qty[-\frac{1}{2m}\nabla^2+V_0-E]\mathcal G(x_0,y_0;x,y)=\delta(x-x_0)\delta(y-y_0),
\end{equation}
where $x_0>0$. The solution is given by
\begin{align}
    \mathcal G(x_0,y_0;x,y)&=\int\frac{\dd[2]{k}}{(2\pi)^2}\frac{e^{ik_x(x-x_0)+ik_y(y-y_0)}}{\frac{k_x^2+k_y^2}{2m}+V_0-E-i\epsilon}\nonumber\\
    &=\frac{mi}{2}H^{(1)}_0\qty(\sqrt{2m(E-V_0)}\sqrt{(x-x_0)^2+(y-y_0)^2}),
\end{align}
where $H^{(1)}_\alpha(x)$ is the Hankel function of the first kind. For $E>V_0$, this is an oscillating solution.
From the Kirchhoff integral theorem, we have
\begin{align}
    \psi_{x\geq0}(x_0,y_0)&=\frac{1}{2m}\int_\Sigma\dd{\bf S}\cdot\qty[\mathcal G(x_0,y_0;x,y)\nabla\psi_{x\geq0}(x,y)-\psi_{x\geq0}(x,y)\nabla \mathcal G(x_0,y_0;x,y)]\nonumber\\
    &=\frac{-1}{2m}\int_{-\infty}^\infty\dd{y}e^{ip_yy}\qty[B\mathcal G(x_0,y_0;0,y)-A\partial_x \mathcal G(x_0,y_0;0,y)],
\end{align}
where $\Sigma$ is an arbitrary boundary that encloses $(x_0,y_0)$.

The integrals can be executed easily in the momentum space and we obtain
\begin{align}
    \int_{-\infty}^\infty\dd{y}e^{ip_yy}\mathcal G(x_0,y_0;0,y)&=\frac{m}{\kappa}e^{-\kappa x_0+ip_yy_0},\\
    \int_{-\infty}^\infty\dd{y}e^{ip_yy}\partial_x\mathcal G(x_0,y_0;0,y)&=me^{-\kappa x_0+ip_yy_0},
\end{align}
where
\begin{equation}
    \kappa=\sqrt{2mV_0-p_x^2}.
\end{equation}
A skeptical reader may check these numerically although the convergence is relatively slow.

We obtain
\begin{equation}
    \psi_{x\geq0}(x,y)=\frac12\qty(A-\frac{B}{\kappa})e^{-\kappa x+ik_yy}=Ae^{-\kappa x+ik_yy},
\end{equation}
where we used Eq.~\eqref{eq_bdy_a}. Thus, we successfully reproduced the correct exponent along the $x$-axis, starting from the oscillating solution.

This phenomenon can also be understood as the total reflection. As the phase velocity at $x\geq0$ is faster than that at $x<0$, the refraction angle becomes complex when $p_x^2/(2m)<V_0$. The decaying wave function at $x\geq0$ is then understood as the evanescent waves.

The above discussion is related to the recent argument of the tunneling time \cite{Sokolovski2021}, where they discuss that the tunneling time cannot be defined because quantum tunneling is a result of the interference of the paths that spent different time in the potential barrier. As we have seen, the correct exponent of the wave function emerged from the interference of propagating waves. Since these waves have the same velocity in the barrier, they spend different time in the potential barrier and one cannot tell which represents the tunneling time.
\section{Example of General Two Body System}\label{apx_two_body}
In this appendix, we derive a path integral formula for a general two-body system using a slightly different procedure than in Section \ref{sec_non_separable}.
We consider two particles with the Hamiltonian given by
\begin{equation}
    H=\frac{p_x^2}{2m_x}+\frac{p_y^2}{2m_y}+V_x(x)+V_y(y)+V_{\rm int}(x-y).
\end{equation}

Integrating $p_a (dp_a/dt)$, the local energy conservation law is obtained for each particle as
\begin{align}
    0&=\frac{p_a^2}{2m_a}+V_a(z^a)+\int\dd{t}\dv{z^a}{t}\pdv{V_{\rm int}}{z^a},
\end{align}
where $z^x=x,z^y=y$.
It motivates us to consider the constraints\footnote{The differences from the procedure in the main text are (i) the interaction energy is included in $\eta^a$ and (ii) $\xi^a$ is introduced to eliminate the degrees of freedom of $\eta^a$. These are only to demonstrate another way to obtain the path integral formula and do not affect the final result essentially.},
\begin{equation}
    \mathcal H_a\approx0,~\mathcal H'_a\approx0,~\pi^\eta_a\approx0,
\end{equation}
where
\begin{align}
    \mathcal H_a&=\frac{p_a^2}{2m_a}+\pi^\xi_a\pi^\eta_a+V_a(z^a)-\frac{1}{\kappa}\eta^a,\\
    \mathcal H'_a&=-\frac{p^a}{m_a}\pdv{V_{\rm int}}{z^a}-\frac{1}{\kappa}\pi^\xi_a.
\end{align}
We can easily see that the Poisson brackets of these constraints give $\{\mathcal H_x,\mathcal H_y\}_P\approx0$, $\{\mathcal H_a,\mathcal H'_b\}_P\not\approx0$ and $\{\mathcal H'_x,\mathcal H'_y\}_P\not\approx0$. Thus, only $\mathcal H_x$ and $\mathcal H_y$ are the involutive constraints.

The constraints are solved after the canonical quantization of the variables,
\begin{equation}
    [\hat z^a,\hat p_b]=i\hbar\delta^a_b,~[\hat \xi^a,\hat \pi^\xi_b]=i\hbar\delta^a_b,~[\hat \eta^a,\hat \pi^\eta_b]=i\hbar\delta^a_b.
\end{equation}

We introduce the deformation variable, $s$, and discretize it as $s=s_1,\dots,s_n$. Then, we define the basis at $s=s_J$ as
\begin{equation}
    \ket{X_J}=\ket{x_J,y_J,\xi_J^x,\xi_J^y,\eta_J^x,\eta_J^y},
\end{equation}
and $S_J$ as
\begin{align}
    \braket{X_{J+1}}{X_J}\bra{X_J}&=\exp{\frac{i}{\hbar}S_J(X_{J+1}-X_J;X_{J+1})}\bra{X_J}.
\end{align}
From $\mathcal H_a\approx0$, we have
\begin{align}
    0&=\Tr\qty[\ket{X_{J+1}}\braket{X_{J+1}}{X_J}\bra{X_J}\hat{\mathcal H}_a]\nonumber\\
    &=\frac1{2m_a}\qty(\pdv{S_J}{z^a})^2+\qty(\pdv{S_J}{\xi^a})\qty(\pdv{S_J}{\eta^a})+V_a(z^a_J)-\frac{1}{\kappa}\eta^a_J.\label{eq_apx_two_ham_const}
\end{align}
The characteristic curve is given by
\begin{align}
    \frac{z^a_{J+1}-z^a_J}{\delta s}&=-\frac{\alpha_J^a}{m}\pdv{S_J}{z^a_J},\\
    \frac{\xi^a_{J+1}-\xi^a_J}{\delta s}&=-\alpha_J^a\pdv{S_J}{\eta_J^a},\\
    \frac{\eta^a_{J+1}-\eta^a_J}{\delta s}&=-\alpha_J^a\pdv{S_J}{\xi_J^a}.
\end{align}
From $\hat\pi^\eta_a\approx0$, we have
\begin{equation}
    \xi_{J+1}^a-\xi_J^a=0,\label{eq_apx_xi_const}
\end{equation}
which eliminates the degrees of freedom of $\xi^a$.
From $\mathcal H'_a\approx0$, we get
\begin{align}
    0&=\Tr\qty[\ket{X_{J+1}}\braket{X_{J+1}}{X_J}\bra{X_J}\hat{\mathcal H}'_a]\nonumber\\
    &=\frac{1}{\alpha^a_J}\qty[-\pdv{V_{\rm int}}{z^a}\frac{z_{J+1}^a-z_J^a}{\delta s}-\frac{1}{\kappa}\frac{\eta_{J+1}^a-\eta_J^a}{\delta s}],\label{eq_apx_eta_const}
\end{align}
which determines $\eta_{J+1}^a-\eta_J^a$ and eliminates $\eta^a$.

In the following, we only consider the paths that satisfy Eqs.~\eqref{eq_apx_xi_const} and \eqref{eq_apx_eta_const}. 
We can determine $\alpha^a$ from Eq.~\eqref{eq_apx_two_ham_const} as
\begin{equation}
   (\alpha^a_J)^2=\frac{\sum_a\frac{m_a}{2}\qty(\frac{z^a_{J+1}-z^a_J}{\delta s})^2}{-V_a(z_J^a)+\frac{\eta^a_J}{\kappa}+i\epsilon}.
\end{equation}
Using the above equations, we obtain
\begin{align}
    \braket{X_{J+1}}{X_J}&\simeq1-\frac{i}{\hbar}\sum_{a}\pdv{S_J}{z^a_J}\frac{z^a_{J+1}-z^a_J}{\delta s}\delta s\nonumber\\
    &=1+\frac{2i}{\hbar}\sum_{a}\sqrt{\frac{m_a}{2}\qty(\frac{z_{J+1}^a-z_J^a}{\delta s})^2}\sqrt{-V_a(z^a_J)+\frac{\eta^a_J}{\kappa}}\delta s.
\end{align}

Finally, we obtain the Green's function as
\begin{align}
    G(X_n,X_1)&=\braket{X_n}{X_1}\nonumber\\
    &=\int\mathcal DX\delta_\eta\exp[\frac{i}{\hbar}\sum_a\int_0^1\dd{s}2\sqrt{\frac{m_a}{2}\qty(\dv{z^a}{s})^2}\sqrt{-V_a(z^a)+\frac{\eta^a}{\kappa}}],
\end{align}
where the path integral, $\int\mathcal DX\delta_\eta$, is executed over the paths that satisfy
\begin{align}
    \dv{\eta^a}{s}&=-\kappa\pdv{V_{\rm int}}{z^a}\dv{z^a}{s}.
\end{align}
Here, we have rescaled $s$ so that $s\in[0,1]$.
Unlike the other examples, we have
\begin{equation}
    2\sqrt{\frac{m_a}{2}\qty(\dv{z^a}{s})^2}\sqrt{-V_a(z^a)+\frac{\eta^a}{\kappa}}=\alpha\qty[\frac{m_a}{2}\qty(\frac{1}{\alpha}\dv{z^a}{s})^2-V_a(z^a)+\frac{\eta^a}{\kappa}],
\end{equation}
which looks different from the naive Lagrangian. It is because $\eta^a$ contains the energy of $V_{\rm int}$ and it does not appear explicitly.
\section{Spherical Symmetry}\label{apx_o3}
It has not been shown that the most probable path is $O(3)$-symmetric for the polychronic tunneling. However, it becomes numerically much easier if we assume the $O(3)$-symmetry. In addition, the gravitational effect is not important at a scale much below the Planck scale.
Thus, in this appendix, we give the formulas with the $O(3)$-symmetry in the decoupling limit of gravity\footnote{We can also include gravity, see \cite{Oshita:2021aux} for the formulas.}.
We utilize the results in Section \ref{sec_gravity} and give only the relevant formulas.

We take $\bfx=(r,\theta,\varphi)$ with the general\footnote{The momentum constraints for $\theta$ and $\varphi$ determine the metric to be this form.} $O(3)$-symmetric metric,
\begin{equation}
    h_{ij}\dd{x^i}\dd{x^j}=(1+A(s,r))\dd{r^2}+(1-B(s,r))r^2(\dd{\theta^2}+\sin^2\theta\dd{\varphi^2}).
\end{equation}
In the decoupling limit of gravity, we obtain $\alpha=\sqrt{\mathcal K}/\sqrt{-\mathcal V}$ with
\begin{align}
    \mathcal K&=\frac{1}{2}(\partial_s\phi)^2+\order{\kappa},\\
    \mathcal V&=V(\phi)+\frac{1}{2}(\partial_r\phi)^2-\frac{1}{\kappa r^2}\partial_rr^2\qty[\frac{A}{r}+\frac{1}{r}\partial_rrB]+\order{\kappa}.
\end{align}
The momentum constraint for $r$ gives
\begin{equation}
    \partial_s\qty[\frac{A}{r}+\frac{1}{r}\partial_rrB]=\kappa(\partial_s\phi)(\partial_r\phi)+\frac{\partial_r\alpha}{\alpha}(\partial_sB).
\end{equation}

We find the Schwarzschild coordinates, {\it i.e. $B=0$}, is the easiest choice since it has a natural geometric interpretation of the surface area and there is no term proportional to $\partial_r\alpha$. 
The Green's function is then given by
\begin{equation}
    G(\Phi_n,\Phi_1)=\int\mathcal D\phi\mathcal DA\delta_A\exp\qty[\frac{8\pi i}{\hbar}\int_0^1\dd{s}\int\dd{r}r^2\sqrt{\mathcal K(s,\bfx)}\sqrt{-\mathcal V(s,\bfx)}].
\end{equation}
Here, $\mathcal DA\delta_A$ is the path integral over the paths satisfying
\begin{equation}
    \partial_s\frac{A}{r}=\kappa(\partial_s\phi)(\partial_r\phi),
\end{equation}
and
\begin{align}
    \mathcal K&=\frac{1}{2}(\partial_s\phi)^2,\\
    \mathcal V&=V(\phi)+\frac{1}{2}(\partial_r\phi)^2-\frac{1}{\kappa r^2}\partial_rrA.
\end{align}

\bibliographystyle{apsrev4-1}
\bibliography{path}
\end{document}